\documentclass[12pt,a4paper]{elsart}

\usepackage[dvips]{graphicx,color}
\usepackage{amsmath}

\newcommand{\und}{\underline}
\newcommand{\D}{{\rm d}}

\renewcommand{\vec}[1]{\underline{#1}}
\newcommand{\bo}{\mathcal{L}^{\dagger}}

\begin{document}

\begin{frontmatter}

\title{Exact Solution of the Multi-Allelic Diffusion Model}

\author[mcr]{G.\ Baxter},
\author[ed]{R.\ A.\ Blythe}, 
\author[mcr]{A.\ J.\ McKane}

\address[mcr]{Theory Group, School of Physics and Astronomy,
University of Manchester, Manchester M13 9PL, UK}

\address[ed]{School of Physics, University of Edinburgh, Mayfield Road, \\
Edinburgh EH9 3JZ, UK}

\begin{abstract}
We give an exact solution to the Kolmogorov equation describing
genetic drift for an arbitrary number of alleles at a given
locus. This is achieved by finding a change of variable which makes
the equation separable, and therefore reduces the problem with an
arbitrary number of alleles to the solution of a set of equations that
are essentially no more complicated than that found in the two-allele
case. The same change of variable also renders the Kolmogorov equation
with the effect of mutations added separable, as long as the mutation
matrix has equal entries in each row. Thus this case can also be
solved exactly for an arbitrary number of alleles. The general
solution, which is in the form of a probability distribution, is in
agreement with the previously known results---which were for the
cases of two and three alleles only. Results are also given for a wide
range of other quantities of interest, such as the probabilities of
extinction of various numbers of alleles, mean times to these
extinctions, and the means and variances of the allele frequencies. To
aid dissemination, these results are presented in two stages: first of
all they are given without derivations and too much mathematical
detail, and then subsequently derivations and a more technical
discussion are provided.
\end{abstract}


\end{frontmatter}

\section{Introduction}
\label{intro}

Although probabilistic ideas are at the heart of genetics, and have
been used since the earliest days of the subject, it was in the study
of genetic drift in the 1920s and 1930s that the notion of stochastic
processes first played a major part in the theory of genetics. In the
simplest case, a diploid population of a fixed size, $N$, was
considered, and attention was focused on two alleles at a particular
locus. If selective differences between the two alleles, as well as
the chances of mutation, are ignored, then the changes in the allele
frequencies are purely due to genetic drift. Assuming discrete
non-overlapping generations, one can ask: what is the probability that
$n$ of the $2N$ genes in the population alive at time $t$ are of a
given type?  This formulation of genetic drift is usually referred to
as the Fisher-Wright model, being used implicitly by
Fisher~\cite{fis30}, and explicitly by Wright~\cite{wri31}.

Although Fisher and Wright did not use this terminology, the
stochastic process they defined through this model is a Markov chain,
since the probabilities of genetic change from one generation to the
next do not depend on the changes made in previous
generations. However the use of Markov chains becomes cumbersome when
the effects of mutation and selection are introduced, and for this
reason there was a move away from the description of the process in
terms of a discrete number of alleles $n=0,1\ldots,2N$, and discrete
generations, to a process of diffusion where the fraction of alleles
of one type is a real random variable $x(t)$ in the interval $[0,1]$
and time is continuous. This type of model actually predates the
Markov chain description~\cite{fis22}, and was further studied by
Wright~\cite{wri45} and Kimura~\cite{kim55a}. It is this formulation
that will concern us here, specifically we will be interested in
neutral evolution---that is, the dynamics of a randomly mating
population which may, or may not, be subject to genetic mutation.

Our motivation for the work presented here lies in the study of an
evolutionary model of language
change~\cite{hul88,hul01,cro05,cro02}. As a first step in the
mathematical formulation of this model, we developed a ``drift model''
of language change \cite{us}, which turns out to be identical to the
diffusion models of population genetics discussed above.  Particularly
relevant in this application to language change is the situation where
the number of variants (the linguistic counterpart to alleles) may be
large.  A survey of the population genetics literature revealed a
considerable amount of work on the diffusion equation for two
alleles~\cite{cro70}, some on three alleles~\cite{kim56} and very
little on the general case of an arbitrary number of
alleles~\cite{kim55b}.  Much of the work originated with Kimura in the
1950s, and developments since then appear scattered throughout the
literature.

In this work we fill a considerable gap in the literature by giving
for the first time an analytic solution of the diffusion equation,
with and without mutation, in the general case of $M$ alleles.  Given
the rather scattered nature and---occasionally---the apparent
obscurity of other results relating to genetic diffusion at a single
locus, we take the opportunity here to present them in a systematic
fashion.  In writing this paper we also have a third aim, namely to
present this information in such a way that is accessible to
geneticists who may not wish to work through a great deal of
mathematical analysis. We will shortly describe the plan of the paper
which we hope will allow us to accomplish this.

After Kimura's pioneering work in the 1950s, there was further work by him
\cite{kim62,kim64,kim69,kim83} and others \cite{lit75b,lit78a,not81,mar81}
on single-locus models, but interest naturally moved to models which involved 
multiple loci and the interactions between them 
\cite{kar75,fel75,fra77,lit78b,gol83,lew88}. Despite the continuing development
of population genetics, very little has been achieved by way of exact solutions
for single-locus models since Kimura \cite{kim56} and Littler 
\cite{lit75b,lit78a}. Occasional work has appeared, both of an analytic nature
\cite{zen89,iiz96,rat01,wan04} and involving numerical simulations 
\cite{gil00,pan01,zia02,che03}, although single locus diffusion models 
continue to receive proportionately less attention.

A number of other factors have been responsible for the relative
neglect of this particular area of population genetics. One of the
main ones is that the focus has shifted to experimental investigations
and there has been the consequential realisation that reality is very
much more complex than what simple models allow.  However, it is also
the case that the theory is perceived as mathematically
difficult.  Even Kimura~\cite{kim56} states when studying the extension
to three alleles that for the partial differential equation describing
diffusion ``the general case of an arbitrary number of alleles can be
solved [by separation of variables]. However, additional techniques
will be needed to make the mathematical manipulations manageable''. We
will show here that this is not the case: in fact it is no harder to
solve the case of the diffusion of a general number of alleles than
the three-allele case.

We believe that, while our motivation for this work originates with
models of language change, geneticists will also find the results we
obtain useful.  The diffusion theory is described in several standard
texts on population genetics~\cite{rou79,har00,kim64b,ewe79} and
general consequences of the results previously found in the form of
simple rules are widely known and utilised. The problem, as we have
already indicated, is the mathematical nature of the theory. The
diffusion approximation in the case without mutation results in a partial 
differential equation which is of the type which is used to describe
the diffusion of heat in a metal or dye in a fluid~\cite{har00}. The
inclusion of mutation corresponds in these physical problems to a
deterministic force which biases the diffusion in one direction rather
than another. For example, if the particles diffusing were charged,
then this deterministic force could arise from an electric field being
imposed on the system. Not surprisingly, diffusion equations of this
kind have been extensively studied by physicists who call them
Fokker-Planck equations~\cite{ris89,gar04}. Here we will use the
nomenclature used in the genetics literature where they are called
Kolmogorov equations. There are other confusing changes in
nomenclature across disciplines: for example, the term representing
the deterministic motion discussed above in the Kolmogorov equation is
called the drift in the physics literature, but in the genetics
literature it is the diffusive term which is called the drift.  In
order to avoid confusion we shall now drop the further use of the term
``drift'' in this work.

The outline of the paper is as follows. In section \ref{overview} we
will give on overview of our results which is devoid of proofs and too
much mathematical detail, so that it may be read by those who simply
wish to have a reasonably non-technical summary of the results of the
paper. It will also serve as a general introduction to those who wish
to go on and absorb the more technical sections of the paper. These
technical sections start with section \ref{soln}, where the general
solution of the Kolmogorov equation for an arbitrary number of
alleles, with and without mutations, is derived. In section
\ref{other} we give details of the calculation of various quantities
of interest that are given briefly at the end of section
\ref{overview}. We end in section \ref{discuss} with an overview of
the paper's main results and point out some remaining unsolved
cases.  Further technical material has been relegated to three
appendices so as not to interrupt the flow of the arguments in the
main text.

\section{Overview of the main results}
\label{overview}
In this section we will describe the problems we investigate, explaining our 
methods of analysing them and give the results we have obtained without
proof. 

\subsection{Definition of the model}
\label{definition}
The simplest, and most widely studied, version of the problems we will
be investigating in this paper, is where two alleles, $a_1$ and $a_2$,
are segregating in a randomly-mating population of $N$ monoecious
individuals.  We begin with the purely random case in which the only
way that allele frequencies may change is through the random sampling
of gametes in the reproductive process. If $x$ is the frequency of
allele $a_1$ and $1-x$ the frequency of allele $a_2$, we would expect
that the value of $x$ would change through time until eventually one
or other of the alleles would become fixed, \textit{i.e.} $x$ would
take on the value 0 or 1. The mathematical quantity which describes
the process is the conditional probability distribution function (pdf)
$P(x, t|x_{0}, 0)$---the probability that the frequency of the $a_1$
allele at time $t$ is $x$, given the frequency of this allele at $t=0$
was $x_0$.  It satisfies the Kolmogorov
equation~\cite{cro70,rou79,har00}
\begin{equation}
\frac{\partial P}{\partial t} = \frac{\partial^{2} }{\partial x^{2}}
\left[ D(x) P \right] \,, \ \ \ D(x) = \frac{1}{2} x(1-x)\,,
\label{Kol_2}
\end{equation}
in the interval $0 < x < 1$. This is a diffusion equation for the
frequency of the $a_1$ allele, but with a diffusion coefficient,
$D(x)$, which depends on the frequency. The form of the diffusion
coefficient is such that it gets much weaker as $x$ approaches the
boundaries at $x=0$ and $x=1$. However---as we show using a simple
argument in Appendix~A---it is still sufficiently strong that fixation
can occur, but once this has happened there is no possibility of
re-entering the region $0 < x < 1$.  (In Feller's classification
\cite{fel52}, these are exit boundaries).  As we shall see, it will be
necessary to consider the boundaries separately from the interior
region when calculating probabilities.

It should also be noted that in the derivation of Eq.~(\ref{Kol_2}) (see 
Section \ref{soln}), one unit of time is taken to be $2N$ generations, so that
the time between generations is $1/(2N)$. Although this choice is useful in
derivations involving the diffusion approximation, it is frequently more
useful to work with $\tau \equiv 2 N t$, so that the unit of time is that 
between generations. Then Eq.~(\ref{Kol_2}) takes the form 
\begin{equation}
\frac{\partial P}{\partial \tau} = \frac{1}{4N} \frac{\partial^{2} }
{\partial x^{2}}\left[ x(1-x) P \right] \,,
\label{tau_form}
\end{equation}
which is the one that generally appears in textbooks \cite{cro70}.

When $M$ alleles $a_{1},a_{2},\ldots,a_{M}$ are present, these ideas
can be generalised. If $x_{1}, x_{2},\ldots, x_{M}$ are the frequencies 
of the alleles, then there are only $M-1$ independent variables---the
frequency $x_M$ may be replaced by $1-x_{1}- \ldots - x_{M-1}$. The
Kolmogorov equation is now an $(M-1)$-dimensional diffusion
equation~\cite{kim55b}
\begin{equation}
\frac{\partial P}{\partial t} = \sum^{M-1}_{i=1} \sum^{M-1}_{j=1}
\frac{\partial^{2} }{\partial x_{i} \partial x_{j}} 
\left[ D_{ij}(\und{x}) P \right] \,,
\label{Kol_M}
\end{equation}
where the diffusion coefficient is now a matrix:
\begin{equation}
D_{ij} (\und{x}) = \left\{ \begin{array}{ll} 
\frac{1}{2} x_{i} (1-x_{i}) , & \mbox{\ if $i=j$} \\
- \frac{1}{2} x_{i}x_{j} , & \mbox{\ if $i \neq j$\,.} 
\end{array} \right.
\label{diff_M}
\end{equation}

If mutations occurring at constant rates are allowed for, these
Kolmogorov equations acquire an extra term, which involves only first
order derivatives.  In the case of two alleles, if $a_2$ mutates to
$a_1$ at a constant rate given by $m_1$ and $a_1$ mutates to $a_2$ at
a constant rate given by $m_2$, then in the absence of the random
process described above, we would expect the frequency of the $a_1$
allele to change according to the \textit{deterministic} differential
equation $\dot{x} = m_{1} (1-x) - m_{2} x$, where $\dot{x} = {\D
x\over\D t}$. If we include both this deterministic process and the
random diffusion process, the Kolmogorov equation takes the
form~\cite{rou79}:
\begin{equation}
\frac{\partial P}{\partial t} = - \frac{\partial }{\partial x} 
\left[ A(x) P \right] + \frac{\partial^{2} }{\partial x^{2}}
\left[ D(x) P \right]\,,
\label{Kol_2_mut}
\end{equation}
where
\begin{equation}
A(x) = m_{1} (1-x) - m_{2} x\,, \ \ D(x) = \frac{1}{2} x(1-x)\,.
\label{AandD_2}
\end{equation}

Finally, if $M$ alleles are present with mutation, we shall in common
with other authors \cite{tie78b}, only study the situation in which
the rate of mutation of the alleles $a_{j}\,(j \neq i)$ to $a_{i}$
occurs at a constant rate $m_i$, independent of $j$. Then the
deterministic equation would be
\begin{equation}
\dot{x}_{i} = m_{i} (1-x_{i}) - \sum_{j \neq i} m_{j} x_{i} 
= m_{i} - \sum_{j} m_{j} x_{i} \equiv A_{i} (\und{x})\,.
\label{A_M}
\end{equation}
The fact that $A_i(\und{x})$ depends only on $x_i$ is a consequence of
mutation rates depending only on the end product of the mutation.  It
is this feature that allows solution by separation of variables of the
resulting Kolmogorov equation which reads
\begin{equation}
\frac{\partial P}{\partial t} = - \sum^{M-1}_{i=1} 
\frac{\partial }{\partial x_{i}} \left[ A_{i} (\und{x}) P \right] +
\sum^{M-1}_{i=1} \sum^{M-1}_{j=1}
\frac{\partial^{2} }{\partial x_{i} \partial x_{j}} 
\left[ D_{ij}(\und{x}) P \right] \,,
\label{Kol_M_mut}
\end{equation}
where $A_{i} (\und{x})$ and $D_{ij} (\und{x})$ are given by Eqs.~(\ref{A_M}) 
and (\ref{diff_M}) respectively.

It is sometimes useful to use the backward form of the Kolmogorov equation, in 
which the derivatives are with respect to the initial values $x_0$ and $t_0$:
\begin{equation}\label{BKE}
\frac{\partial}{\partial t_0} P(\und{x},t|\und{x}_0,t_0)= - \sum^{M-1}_{i=1}
A_{i} (\und{x}_{0}) \frac{\partial }{\partial x_{i,0}}P -
\sum^{M-1}_{i=1} \sum^{M-1}_{j=1} D_{ij}(\und{x}_{0})
\frac{\partial^{2} }{\partial x_{i,0} \partial x_{j,0}} 
 P  \,\equiv -\bo P\,.
\end{equation}
Since for the processes we will be interested in here, 
$P(\und{x}, t|\und{x}_{0}, t_{0})$ will be a function only of $(t-t_{0})$, the
left-hand side may also be written as $-\partial P/\partial t$, and $t_0$
chosen to be equal to zero. A detailed derivation and explanation of the 
relationship between the forward and backward forms can be found in 
\cite{ris89}.

Note that immigration is commonly introduced \cite{kim64b} by the use of 
$A(x)=m(\bar{x}-x)$ where $\bar{x}$ represents the mean of the population 
where the immigrants come from. This situation can be represented by 
Eq.~(\ref{Kol_2_mut}) simply by setting $m_1=m\bar{x}$ and $m_2=m(1-\bar{x})$. 
When one or more of the mutation parameters is zero, we have irreversible 
mutation, when the final result is fixation to those alleles which have $m_i$ 
not equal to zero. We will not consider this special case here, but see for 
example \cite{wri38,kim85}.

In purely mathematical terms, the purpose of this paper is to obtain
the conditional pdf $P(\und{x}, t|\und{x}_{0}, 0)$ obtained by solving
Eqs.~(\ref{Kol_M}) and (\ref{Kol_M_mut}), subject to the initial
conditions that $x_{i} = x_{i,0}$ at $t=0$ and the appropriate
boundary conditions. As far as we are aware, only the cases $M=2, 3$
without mutation and $M=2$ with mutation have been solved to
date~\cite{cro70}. The key to making progress is to find a change of
variable which makes these equations separable and also to note that
the solution with $n$ alleles is nested within the solution with $n+1$
alleles, so many properties follow by induction on $n$.

\subsection{Change of variables}
\label{u_variables}
While the $M=2$ Kolmogorov equations (\ref{Kol_2}) and (\ref{Kol_2_mut}) can be
solved by separation of variables---writing $P(x, t)=X(x) T(t)$---their
counterparts for general $M$, Eqs.~(\ref{Kol_M}) and (\ref{Kol_M_mut}) cannot. 
The standard way to proceed in such a situation is to look for a transformation
to a coordinate system in which these partial differential equations are 
separable. Kimura~\cite{kim56} gave such a transformation for values of $M$
up to 4 without mutation, but not in general. There is no systematic way to 
discover such transformations, but we have found one which works for all $M$, 
namely:
\begin{equation}
u_i = \frac{x_i}{1 - \sum_{j<i} x_j}\,, \ \ i=1,\ldots, M-1\,,
\label{cofv}
\end{equation}
with the inverse transformation
\begin{equation}
x_i = u_i\,\prod_{j<i} (1-u_j)\,.
\label{cofv_inverse}
\end{equation}
As discussed in section \ref{soln}, this change of variables means that the 
most general form of the Kolmogorov equation we consider, 
Eq.~(\ref{Kol_M_mut}), may be written in diagonal form:
\begin{equation}
\frac{\partial \mathcal{P}}{\partial t} = - \sum^{M-1}_{i=1} 
\frac{\partial }{\partial u_{i}} \left[ \mathcal{A}_{i} (\und{u}) 
\mathcal{P} \right] +\sum^{M-1}_{i=1}\frac{\partial^{2} }{\partial u_{i}^{2}} 
\left[ \mathcal{D}_{i}(\und{u}) \mathcal{P} \right] \,,
\label{Kol_M_u}
\end{equation}
that is, there are no mixed second-order derivatives. For completeness we
give the expressions for the functions $\mathcal{A}_{i} (\und{u})$ and 
$\mathcal{D}_{i} (\und{u})$:
\begin{equation}
\mathcal{A}_{i} (\und{u}) = \frac{ \left\{ m_{i} - 
\left( \sum^{M}_{j=i} m_{j} \right) u_{i} \right\} }{ \prod_{j<i}
(1-u_j)}\,, \ \ 
\mathcal{D}_{i} (\und{u}) = \frac{1}{2} \frac{u_{i} (1-u_i)}
{\prod_{j<i} (1-u_j)}\,.
\label{calA_calD}
\end{equation}
Note that this equation has been derived using the new variables---just
changing variables using (\ref{cofv}) in the original Kolmogorov equation
(\ref{Kol_M_mut}) will not give (\ref{Kol_M_u}), but will instead simply give 
an equation for $P(\und{x}, t|\und{x}_{0}, 0)$ with $x_i$ replaced by 
$u_{i} \prod_{j<i} (1-u_j)$. The new conditional pdf in (\ref{Kol_M_u}),
denoted by $\mathcal{P} (\und{u}, t|\und{u}_{0}, 0)$, is related to 
$P(\und{x}, t|\und{x}_{0}, 0)$ through the Jacobian of the transformation to 
the new variables:
\begin{eqnarray}
\nonumber
\mathcal{P}(\und{u}, t|\und{u}_{0}, 0)\,\D u_{1} \D u_{2} \ldots \D u_{M-1} &=&
P(\und{x}, t|\und{x}_{0}, 0)\,\D x_{1} \D x_{2} \ldots \D x_{M-1} \\
\Rightarrow \ \ \mathcal{P} (\und{u}, t|\und{u}_{0}, 0) &=&
P(\und{x}, t|\und{x}_{0}, 0)\,\frac{\partial (x_{1},\ldots,x_{M-1})}
{\partial(u_{1},\ldots,u_{M-1})}\;.
\label{trans_prob}
\end{eqnarray}
The Jacobian is calculated in Appendix B. When $M=3$ it is simply $(1-u_1)$,
so that
\begin{equation}
\mathcal{P} (u_{1}, u_{2}, t| u_{1,0}, u_{2,0}, 0) = \left( 1 - u_{1} \right) 
P(x_{1}, x_{2}, t| x_{1,0}, x_{2,0}, 0)\;,  
\label{trans_prob_3}
\end{equation}
and in the general case one finds that
\begin{equation}
\mathcal{P} (\und{u}, t|\und{u}_{0}, 0) = \left( 1 - u_{1} \right)^{M-2} 
\left( 1 - u_{2} \right)^{M-3} \ldots \left( 1 - u_{M-2} \right) 
P(\und{x}, t|\und{x}_{0}, 0)\;.
\label{trans_prob_M}
\end{equation}

We may summarise this subsection in the following way. We have given a
change of variables which allows the Kolmogorov equation to be written in
a form which is separable. It is easier to work with the conditional pdf
appropriate to these variables, and then to transform back to the original 
variables of the problem and the original conditional pdf using 
Eqs.~(\ref{cofv}) and (\ref{trans_prob_M}) respectively.

\subsection{Expressions for the conditional probability distribution functions}
\label{cond_pdf}
The details of the solution of the Kolmogorov equation (\ref{Kol_M_u}) are
given in section 3 and Appendix C. Here we will present only the final 
expressions.

\medskip

\underline{(i) $M=2$, no mutations.} For orientation, let us first give the 
expression in the case $M=2$ and where no mutation is present. This is well
known and appears in the standard texts~\cite{cro70}. In this case no change 
of variable is required: $u=x$. Then
\begin{equation}
P(x,t|x_{0},0) = x_{0}(1-x_{0}) \sum^{\infty}_{l=0} 
c_{l} F_{l} (1-2x_0) F_{l} (1-2x) 
\exp\left\{-\frac{1}{2}(l+1)(l+2)t\right\}\;.
\label{soln_2}
\end{equation}
The constant $c_l$ is given by $c_{l} = [(2l+3)(l+2)]/(l+1)$ and the 
$F_{l}$ are polynomials, described below. The function (\ref{soln_2})
satisfies both the initial and boundary conditions and holds for all $x$ in 
the interval $(0,1)$, that is, excluding the boundaries at $x=0$ and $x=1$. 
For not too small values of $t$, the solution is well approximated by 
keeping only the first few terms in $l$. In these cases the polynomials 
have a simple form, since $F_{l} (z)$ is a polynomial in $z$ of order 
$l$. For instance,
\begin{equation}
F_{0} (z) = 1\,, \ \ F_{1} (z) = 2z\,, \ \ 
F_{2} (z) = - \frac{3}{4} \left( 1 - 5z^{2} \right)\,, \ldots \,.
\label{low_order_2}
\end{equation}
The natural variable for these polynomials is $z=1-2x$, so that they are 
defined on the interval $-1 < z < 1$. For general $l$ they are the Jacobi 
polynomials $P^{(1,1)}_{l} (z)$~\cite{abr65}.

\begin{figure}[tb]
\begin{center}
\includegraphics[width=0.45\linewidth]{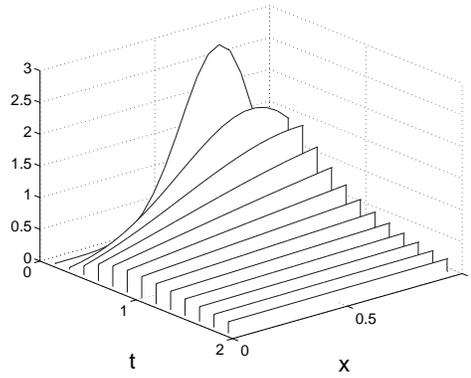}
\end{center}
\caption{\label{neut_2a} Time evolution of the pdf for a biallelic
  system as determined from the analytic solution  with no mutation and
  initial condition $x_0=0.7$.}
\end{figure}

We plot the time development of the solution (\ref{soln_2}) in 
Fig.~\ref{neut_2a}. Beginning from a delta function at the initial point, the 
distribution initially spreads out until the boundaries are reached. The 
distribution is soon nearly flat, and subsides slowly as probability escapes 
to the boundaries. Note that the probability deposited on the boundaries is 
not shown here (but will be discussed later).

\medskip

\underline{(ii) General $M$, no mutations.} For general $M$, the solution 
may be written in the form
\begin{equation}
\mathcal{P} (\und{u}, t|\und{u}_{0}, 0) = w(\und{u}_{0}) \sum_{\lambda} 
\Phi_{\lambda} (\und{u}_{0}) \Phi_{\lambda} (\und{u}) e^{-\lambda t}\,.
\label{soln_M_first}
\end{equation}
as discussed in Section \ref{soln}. The solution is written in terms of the 
$\und{u}$ variables since the problem is separable and so the eigenfunctions 
are separable. The function $w(\und{u})$ is given by
\begin{equation}
w(\und{u}) = \prod^{M-1}_{i=1} u_{i} \left( 1 - u_{i} \right)\,,
\label{explicit_weight}
\end{equation}
and $\lambda$ depends on a set of $M-1$ non-negative integers 
${l_{1},l_{2},\ldots,l_{M-1}}$ according to
\begin{equation}
\lambda = \frac{1}{2} L_{M-1} \left( L_{M-1} + 1 \right)\,, \ \ \ 
L_{i} = \sum^{i}_{j=1} \left( l_{j} + 1 \right)\,.
\label{lambda_and_L}
\end{equation}
This means that the sum over $\lambda$ in Eq.~(\ref{soln_M_first}) is in
fact an $(M-1)-$dimensional sum. 

The property, which we have already remarked upon, that the system
with $M-1$ alleles is nested in that with $M$, manifests itself here
in the fact that the functions $\Phi_{\lambda}$ are very closely
related to the functions which already appear in the two allele
solution (\ref{soln_2}). First of all, since the problem is separable
in this coordinate system, they may be written as
\begin{equation}
\Phi_{\lambda} (u_{1},u_{2},\ldots,u_{M-1}) = \prod^{M-1}_{i=1} 
\psi^{(i)} (u_i)\,.
\label{soln_M_second}
\end{equation}
Each of the factors $\psi^{(i)}$ in this product separately satisfies
the same differential equation.  We shall derive this equation in
Section~\ref{soln}, from which we will learn that $\psi^{(i)}$ depends
only on $u_i$ and on the integers $l_i$ and $L_{i-1}$. Specifically,
\begin{equation}
\psi^{(i)} (u_i) = \left[ c_{l_i} (L_{i-1}) \right]^{1/2} 
\left( 1 - u_{i} \right)^{L_{i-1}} P^{(1,2L_{i-1}+1)}_{l_i} (1-2u_i)\,.
\label{soln_M_third}
\end{equation}
Here the $c_{l_i}$ are the analogue of the $c_l$ that appeared in
Eq.~(\ref{soln_2}), but in this case we have included them in the
function $\Phi_{\lambda}$, so that it satisfies the simple orthogonality
relation
\begin{equation}
\int \D\und{u}\,w(\und{u}) \Phi_{\lambda} (\und{u}) \Phi_{\lambda '} (\und{u})
= \delta_{\lambda\,\lambda '}\,. 
\label{simple_ortho}
\end{equation}
The explicit form of these constants is
\begin{equation}
c_{l_i} (L_{i-1}) = \frac{(2l_{i}+3+2L_{i-1})(l_{i}+2+2L_{i-1})}{(l_{i}+1)}\,.
\label{soln_M_fourth}
\end{equation}
The $P^{(1,\beta)} (1-2u)$, now with $\beta=2L_{i-1}+1$, are again Jacobi 
Polynomials~\cite{abr65}.

\begin{figure}[tb]
\begin{center}
\includegraphics[width=0.95\linewidth]{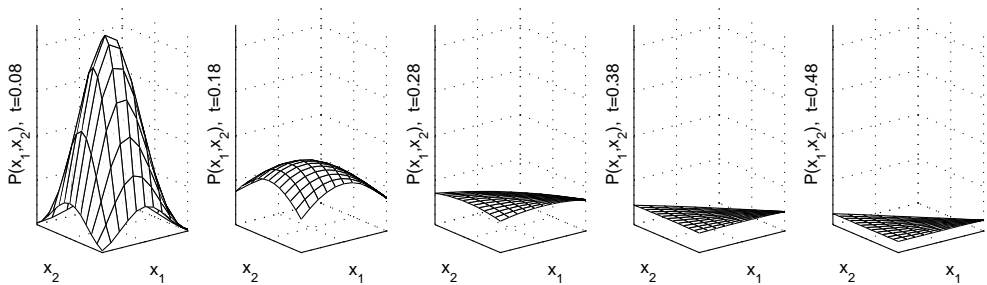}
\end{center}
\caption{\label{neut_3a} Time evolution of the pdf for a triallelic
system as determined from the analytic solution  with no mutation and
initial condition $x_{1,0}=x_{2,0}=0.33$.}
\end{figure}

The general solution for $M>2$ has previously only been found for the
case $M=3$. We have checked that our solution agrees with the
published result~\cite{kim56} in this case, although the labels on the
alleles are permuted between Kimura's solution and ours ($x_{1}
\leftrightarrow x_{3}$) due to a different change of variable being
made. We plot this solution for a triallelic system in Fig.~\ref{neut_3a} at 
several different times. Just as in the two allele system, the distribution 
initially spreads out and forms a bell shape, which quickly collapses, and 
becomes nearly flat, then subsides slowly as probability escapes to the 
boundaries. Note that the probability deposited on the boundaries is not 
shown here. Notice also the triangular shape of the region over which the 
distribution is supported.

\medskip

\underline{(iii) $M=2$, with mutations.}  When mutation is present,
the solution is again well known~\cite{cro70} for the $M=2$
case. Again, no change of variable is required: $u=x$. Then
\begin{multline}
P(x,t|x_0,0) = \\\sum_{l=0}^{\infty} c_l x^{\alpha} (1-x)^{\beta}
P_l^{(\alpha,\beta)}(1-2x_0) P_l^{(\alpha,\beta)}(1-2x)
\exp \left\{ -\frac{1}{2} l(2R+l-1)t \right\} \;.
\label{soln_2_m}
\end{multline}
The constant $c_l$ is
given by
\[c_{l} =
\frac{\Gamma(2R+l-1)\Gamma(l+1)}{\Gamma(2m_1+l)\Gamma(2m_2+l)}
(2R+2l-1)
\]
while $\alpha=2m_1-1$, $\beta=2m_2-1$ and the parameter $R=m_1+m_2$.

The $P^{(\alpha,\beta)}_{l}$ are Jacobi
polynomials~\cite{abr65}. Unlike the case without mutations,
$\alpha\neq\beta$, and they cannot be written in terms of simpler 
polynomials as in case (i). The function (\ref{soln_2_m}) satisfies both 
the initial and boundary conditions and holds for all $x$ in the interval
$[0,1]$, that is, now including the boundaries at $x=0$ and
$x=1$. This difference is due to the nature of the boundary conditions
in the mutation case, which will be described in detail in Section 3.

For not too small values of $t$, the solution is well approximated by
keeping only the first few terms in $l$. In these cases the
polynomials do have a relatively simple form, since
$P^{(\alpha,\beta)}_{l} (z)$ is a polynomial in $z$ of order $l$. For
instance,
\begin{equation}
P^{(\alpha,\beta)}_{0} (z) = 1\,, \ \ P^{(\alpha,\beta)}_{1} (z) =
\frac{1}{2}[(\alpha+\beta+2)z+\alpha-\beta]\,, \ldots \,.
\label{low_order_2_m}
\end{equation}
Also note that once again the natural variable for these polynomials
is $z=1-2x$, so that they are defined on the interval $-1 \le z \le
1$.  In terms of the allele $a_1$ frequency $x$ and the mutation rates
$m_1$ and $m_2$, these polynomials take the form
\begin{equation}
P^{(2m_1-1,2m_2-1)}_{0} (1-2x) = 1\,, \ \ P^{(2m_1-1,2m_2-1)}_{1} (1-2x) =
2[m_1 - R x]\,, \ldots \,.
\label{low_order_2_mx}
\end{equation}

\begin{figure}[tb]
\begin{center}
\includegraphics[width=0.45\linewidth]{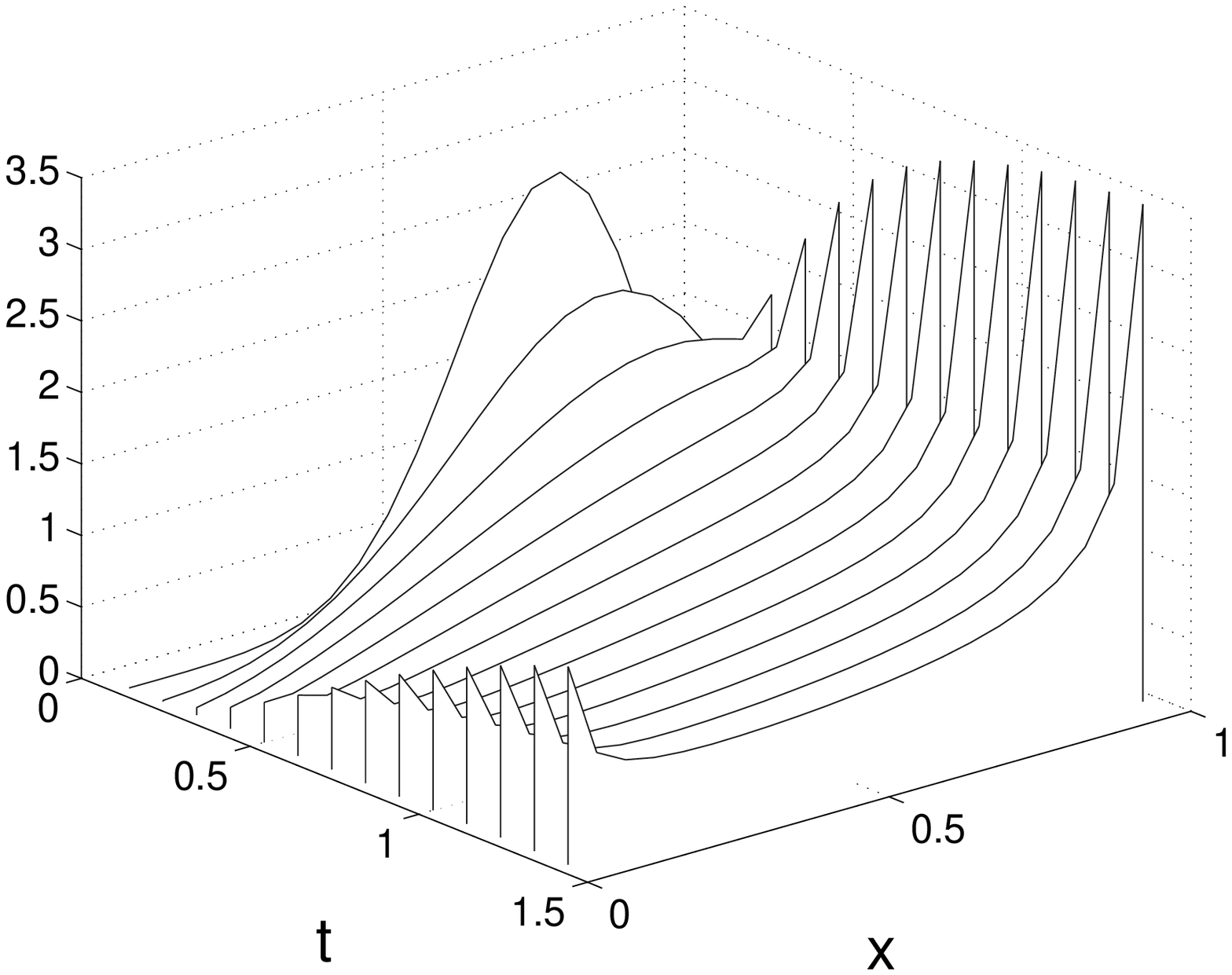}
\hspace{0.05\linewidth}
\includegraphics[width=0.45\linewidth]{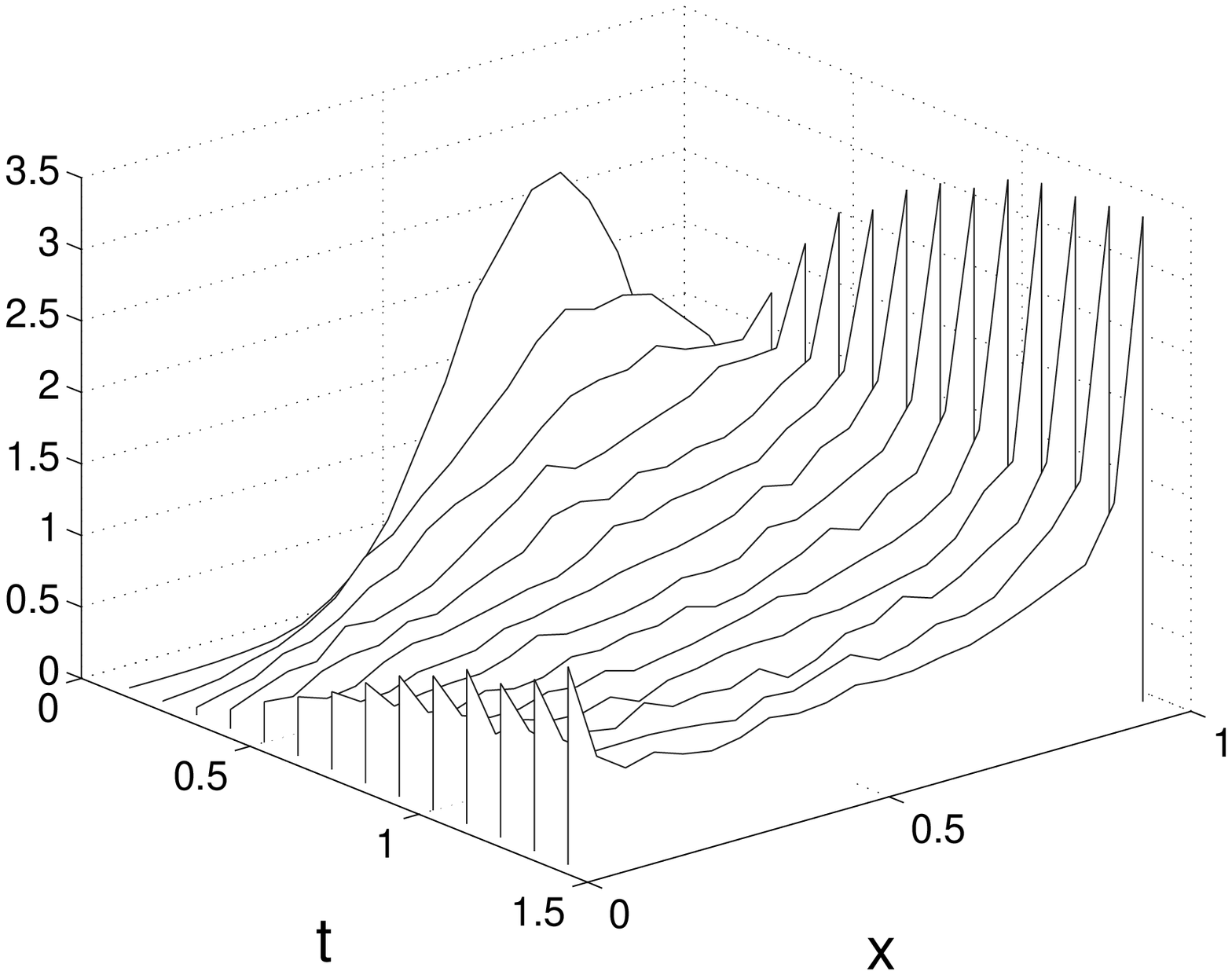}
\end{center}
\caption{\label{mut_2a} Time evolution of the pdf for a biallelic
  system as determined from the analytic solution (left) and Monte
  Carlo simulation (right) with mutation parameters $m_1=m_2=0.2$ and
  initial condition $x_0=0.7$.}
\end{figure}

\begin{figure}[tb]
\begin{center}
\includegraphics[width=0.45\linewidth]{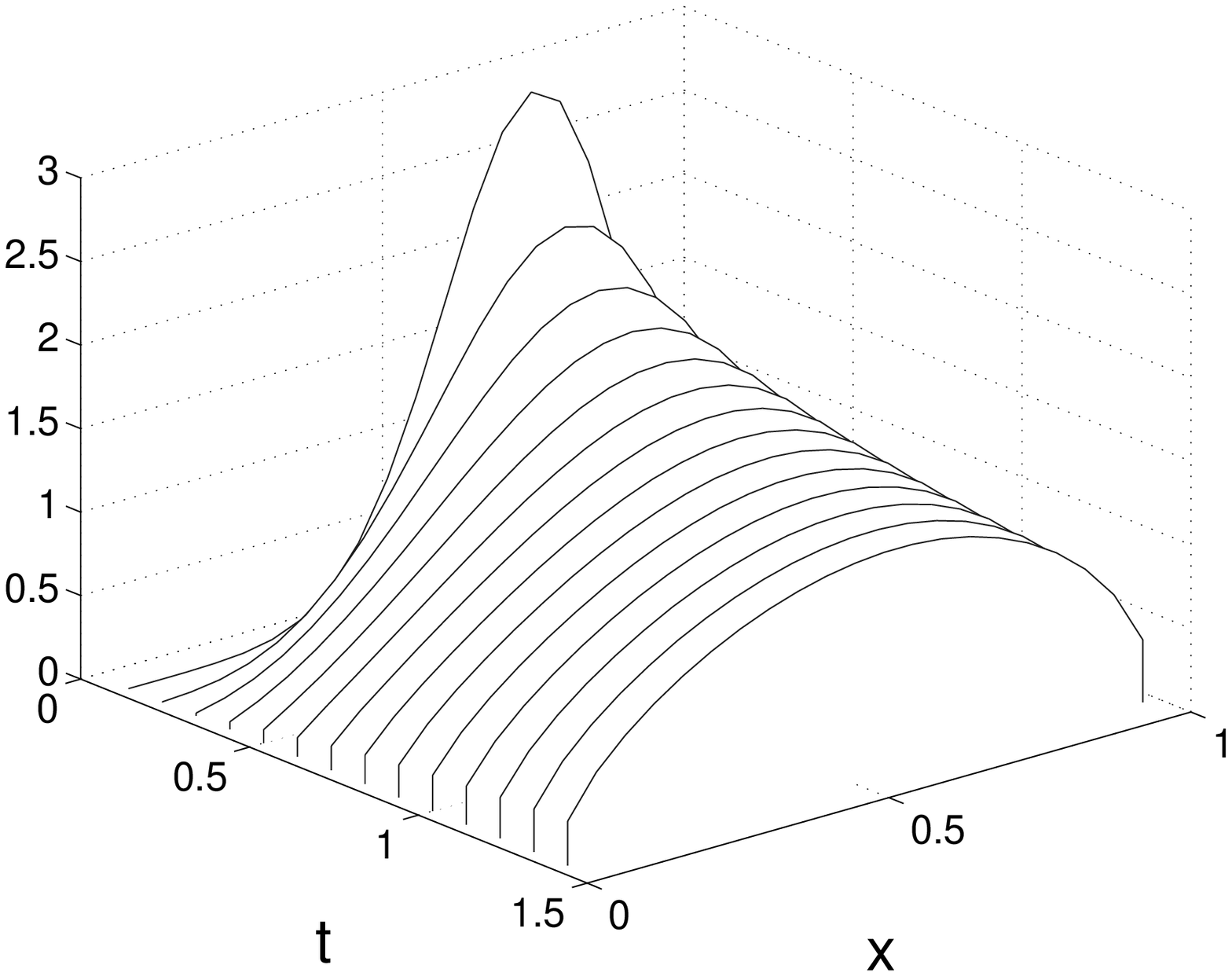}
\hspace{0.05\linewidth}
\includegraphics[width=0.45\linewidth]{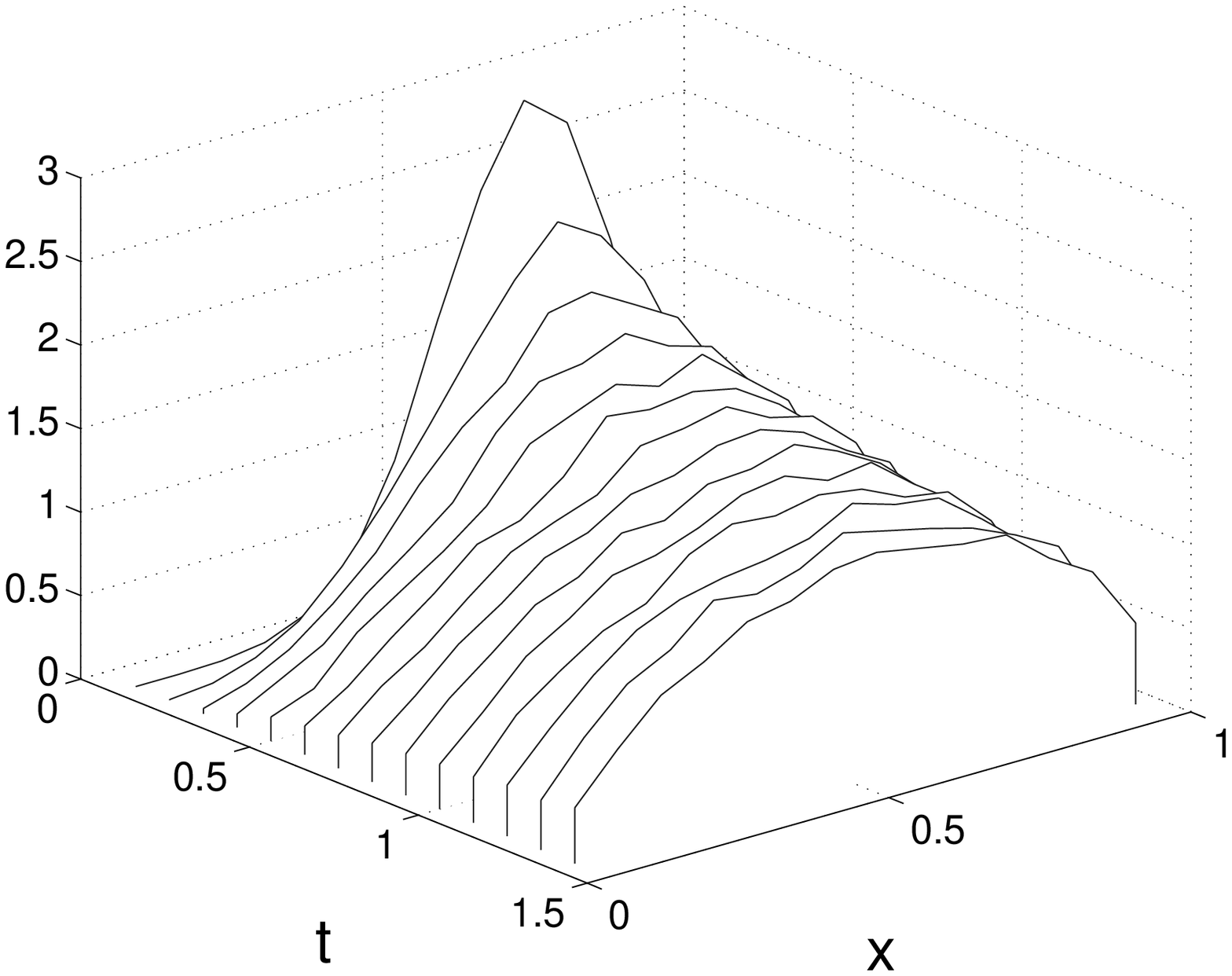}
\end{center}
\caption{\label{mut_2b} Time evolution of the pdf for a biallelic
  systems as determined from the analytic solution (left) and Monte
  Carlo simulation (right) with mutation parameters $m_1=m_2=0.8$ and
  initial condition $x_0=0.7$.}
\end{figure}

For concreteness, we plot in Figs.~\ref{mut_2a} and \ref{mut_2b} the
time evolution pdfs for biallelic ($M=2$) systems from a fixed initial
condition of $x_0=0.7$, i.e., that 70\% of the population have one
particular allele type. In the former case, the mutation rates are
low ($m_1=m_2=0.2$) and one finds a high probability of one allele
dominating the population.  Conversely, in Fig.~\ref{mut_2b} which
has higher mutation rates $m_1=m_2=0.8$, the most probable outcome is
for both alleles to coexist in the population, as indicated by the
peak of the distribution occurring for some value of $x$ far from
either boundary.  In each of the figures, we compare the distribution
obtained from a Monte Carlo simulation of the population dynamics
against the analytic result (\ref{soln_2_m}) with good agreement.

\medskip

\underline{(iv) General $M$, with mutations.} For general $M$, the
solution may be written in the form
\begin{equation}
\mathcal{P} (\und{u}, t|\und{u}_{0}, 0) = \sum_{\lambda} 
\Theta_{\lambda} (\und{u}_{0}) \Phi_{\lambda} (\und{u}) e^{-\lambda t}\,.
\label{soln_M_first_m}
\end{equation}
as discussed in Section \ref{soln}. Here $\Theta_{\lambda} (\und{u})$ is
the left-eigenfunction of the Kolmogorov equation ($\Phi_{\lambda} (\und{u})$
being the right-eigenfunction, as usual). Again, $\lambda$ depends on a set
of $M-1$ non-negative integers ${l_{1},l_{2},\ldots,l_{M-1}}$, with
\begin{equation}
\lambda = \frac{1}{2} \tilde{L}_{M-1} 
\left(2R_{1}+ \tilde{L}_{M-1} - 1 \right)\,, 
\ \ \ \tilde{L}_{i} = \sum^{i}_{j=1}  l_{j} \,, \ \ \ R_{1} = \sum^{M}_{i=1} 
m_{i}\,.
\label{lambda_and_L_m}
\end{equation} 
The left-eigenfunction $\Theta_{\lambda}(\und{u}_0)$ is equal to
$\Phi(\und{u}_0)/P_{\rm st} (\und{u}_{0})$~\cite{ris89}, where 
$P_{st} (\und{u}_{0})$ is the stationary pdf. This is equivalent to the
form (\ref{soln_M_first}), appropriate when there are no mutations, since 
$(P_{st} (\und{u}_{0}))^{-1}$ is what is there called the weight function (up 
to normalisation). Since the problem is separable in this coordinate system, 
$\Phi$ may again be written as
\begin{equation}
\Phi_{\lambda} (u_{1},u_{2},\ldots,u_{M-1}) = \prod^{M-1}_{i=1} 
\psi^{(i)} (u_i)\,.
\label{soln_M_second_m}
\end{equation}
Once again each of the factors $\psi^{(i)}$ is separately a solution
of a single differential equation, and depends on $u_i$, the integers
$l_{M-i}$ and $\tilde{L}_{M-i-1}$ and the mutation rates $m_i$. Specifically,
\begin{equation}
\psi^{(i)} (u_i) = c^{\rm Right}_{l_{M-i}} 
u_{i}^{\alpha_i}\left( 1 - u_{i} \right)^{\beta_i-\tilde{L}_{M-i-1}}
 P^{(\alpha_i,\beta_i)}_{l_{M-i}} (1-2u_i)\,,
\label{soln_M_third_m}
\end{equation}
where
\begin{eqnarray}
c^{\rm Right}_{l_{M-i}} &=& \sqrt{\frac{\Gamma(2R_i)} 
{\Gamma(2m_i) \Gamma(2R_{i+1})}} \nonumber \\
&\times& \sqrt{\frac{(2l_{M-i}+\alpha_{i}+\beta_{i}+1) \Gamma(l_{M-i}+1)
\Gamma(l_{M-i}+\alpha_{i} +\beta_{i}+1)}
{\Gamma(l_{M-i}+\alpha_{i}+1)\Gamma(l_{M-i}+\beta_{i}+1)}}
\label{soln_M_fourth_m}\\
\alpha_i &=& 2m_i-1\\
\beta_i &=& 2R_{i+1}+2\tilde{L}_{M-i-1}-1\\
R_i &=& \sum_{j=i}^{M} m_j \;.
\end{eqnarray}

In a similar way, $\Theta_{\lambda}$ may be written as
\begin{equation}
\Theta_{\lambda} (u_{1},u_{2},\ldots,u_{M-1}) = \prod^{M-1}_{i=1} 
\theta^{(i)} (u_i)\,.
\label{soln_M_fifth_m}
\end{equation}
Each $\theta^{(i)}$ depends on $u_i$ and on the integer
$\tilde{L}_{M-i-1}$ and the mutation constants $R_i$ and $m_i$. Specifically,
\begin{equation}
\theta^{(i)} (u_i) = c^{\rm Left}_{l_{M-i}} 
\left( 1 - u_{i} \right)^{\tilde{L}_{M-i-1}} 
P^{(\alpha_i,\beta_i)}_{l_{M-i}} (1-2u_i)\,,
\label{soln_M_sixth_m}
\end{equation}
where
\begin{eqnarray}
c^{\rm Left}_{l_{M-i}} &=& \sqrt{\frac{\Gamma(2m_i) 
\Gamma(2R_{i+1})}{\Gamma(2R_i)}} \nonumber \\
&\times& \sqrt{\frac{(2l_{M-i}+\alpha_{i}+\beta_{i}+1)) \Gamma(l_{M-i}+1)
\Gamma(l_{M-i}+\alpha_{i} +\beta_{i}+1)}
{\Gamma(l_{M-i}+\alpha_{i}+1)\Gamma(l_{M-i}+\beta_{i}+1)}}\,.
\label{soln_M_left_m}
\end{eqnarray}
The general solution has previously only been found for the case 
$M=2$. Our solution agrees with the published result~\cite{kim56} in this case.

\subsection{Evolution of the population}
\label{calculations}

We can calculate the time evolution of various statistics of the
population.  As previously, we present here a summary of the results
for ease of reference, deferring detailed calculations to
Section~\ref{other}.

\subsubsection{Fixation and extinction}
\label{fixation}

In the absence of mutation, alleles can become extinct since, once
they have vanished from the population, there is no mechanism by which
they can be replaced.  As we have already mentioned, the solution we
have derived (\ref{soln_M_first}) gives the pdf for all states in the
interior. The pdf on a boundary is also given by this equation, so long
as this boundary is not the point of fixation of the final allele, since
it corresponds to the solution with $M$ reduced by 1. So, for example, 
one can read off the pdf in, say, the two-dimensional subspace where alleles 
$a_1$ and $a_2$ coexist and all others present in the initial population 
have become extinct since it is just the $M=2$ solution. Kimura \cite{kim55b}
finds the large-time form for this quantity given an initial condition
comprising $M=3$ alleles using an approach involving moments of the 
distribution which, as we shall show below, can be found directly from the 
Kolmogorov equation without recourse to the full solution we have obtained 
here.

It remains then to determine the fixation probabilities that are
excluded from the pdf we have derived.  Using a simple argument
(outlined below) one can find the probability that a particular allele
has fixed by time $t$ by reduction to an equivalent two-allele
description, the properties of which are well-known \cite{kim55b}.
Using the backward Kolmogorov equation (\ref{BKE}), one can find
further quantities relating to extinction events.  Although some of the
formul\ae\ we quote have appeared in the literature before
\cite{lit75b,lit78b} they seem not to be widely known and some were
stated without proof. We therefore feel there is some value in reviving 
them here and giving what is hopefully a clear derivation in this 
section or in section \ref{other}.

\medskip

\underline{Fixation.}  Define $f_1(x_0,t)$ as the probability that
allele $a_1$, which had initial frequency $x_0$ in a two allele
system, has become fixed by time $t$.  This is given by \cite{kim55b}
\begin{equation}
\label{f_1}
f_1(x_0,t) = x_0 - \frac{1}{2} \sum^{\infty}_{l=0} (-1)^{l} \left[
P_{l} (1-2x_0)- P_{l+2} (1-2x_0) \right] e^{- (l+1)(l+2) t /2 }\,.
\end{equation}
Here $P_l(z)$ is a Legendre polynomial which is a particular case of
the Jacobi polynomials: $P_{l}(z)=P_l^{(0,0)}(z)$.  Conversely, the 
probability that $a_1$ has become extinct by time $t$ is equal to that 
of $a_2$ become fixed, viz,
\begin{equation}
f_2(x_{0},t) = (1-x_{0}) - \frac{1}{2} \sum^{\infty}_{l=0} 
\left[ P_{l} (1-2x_{0})- 
P_{l+2} (1-2x_{0}) \right] e^{-(l+1)(l+2)t/2} \;.
\end{equation}

Now, consider an initial condition with $M>2$ alleles, divided into
two groups $X$ and $Y$.  Let now $x$ be the sum of the frequencies of
$X$ alleles. We may think of the set of alleles $X$ as a single allele in a 
two allele system (for instance, $a_1$ in the above discussion) and the set 
$Y$ as the other allele. Then the probability that \emph{all} of the $Y$ 
alleles (and possibly some of the $X$ alleles) have become extinct by time 
$t$ is simply $f_1(x_0,t)$ where $x_0$ is the initial combined frequency of 
$X$ alleles.  Similarly, the probability that all of the $X$ alleles are
extinct at time $t$, leaving some combination of $Y$ alleles is
$f_2(x_0,t)$. Arguments of this kind, where a set of alleles is identified 
with a single allele in the $M=2$ problem, can be frequently employed to
obtain results for the general case of $M$ alleles in terms of those already
explicitly calculated for $M=2$.

\medskip

\underline{Coexistence probability.}  By combining the above reduction
to an equivalent two-allele problem with combinatorial arguments, the
probability that exactly $r$ alleles coexist at time $t$ can be
calculated \cite{lit75b}.  The result is
\begin{equation}\label{prob_exactly_r}
\Omega_r(\und{x}_0,t)=\sum_{s=1}^r(-1)^{r-s} \binom{M-s}{r-s} \sum
f_1(x_{i_1,0} + \ldots +x_{i_s,0},t)\,,
\end{equation}
where $r$ can take any value from $1$ up to $M$, and the second
summation is over all subsets of size $s$ drawn from the $M$ alleles.
It is a simple matter to use this formula to calculate the mean number
of coexisting alleles at time $t$.

\begin{figure}
\begin{center}
\includegraphics[width=0.45\linewidth]{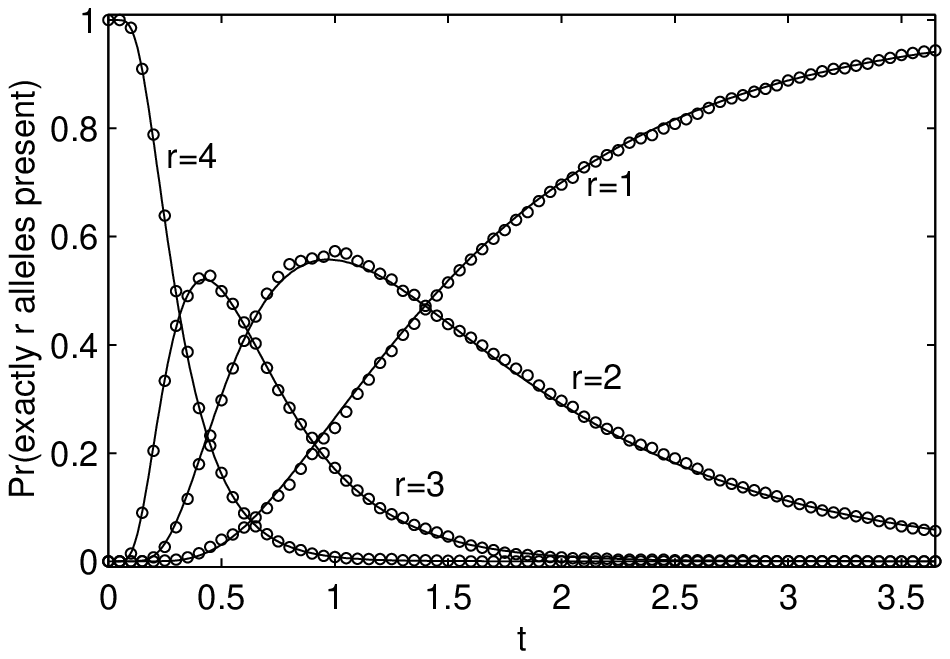}
\hspace{0.05\linewidth}
\includegraphics[width=0.45\linewidth]{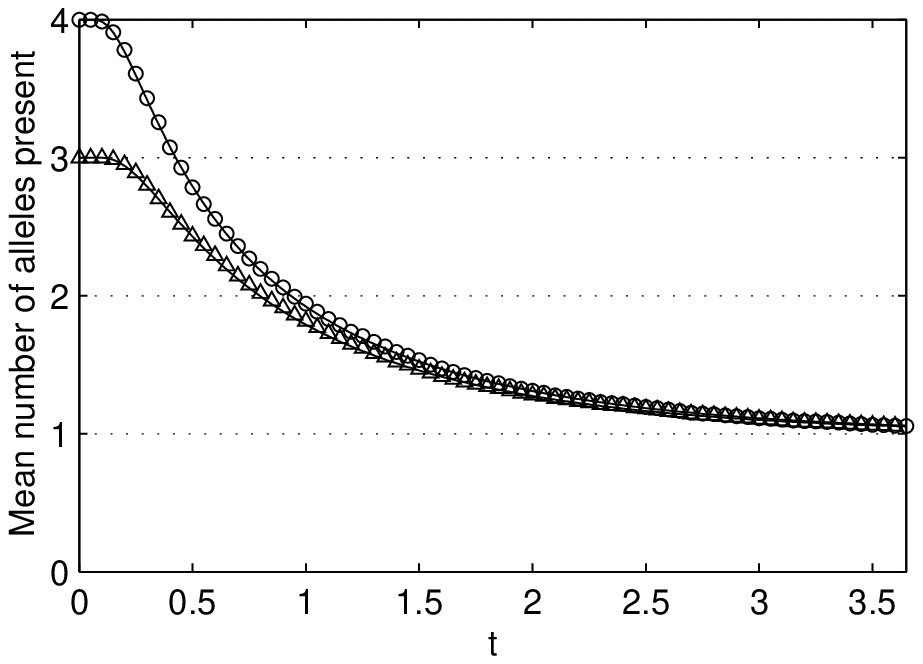}
\caption{Probabilities that exactly $r$ alleles are present as a
function of time for $r=1,2,3,4$ in a system with 4 alleles initially.
Expected number of alleles present as a function of time for systems
with 3 and 4 alleles initially present (right).  Symbols are Monte
Carlo simulation data, solid curves are analytic
calculations.}\label{r_alleles}\label{num_alleles}
\end{center}
\end{figure}

In Fig.~\ref{r_alleles}, we compare the time-dependent probabilities
for $1\le r\le 4$ alleles to be present in a system initially
comprising $M=4$ alleles given by the exact formula
(\ref{prob_exactly_r}) and as obtained from a Monte-Carlo simulation 
with the allele frequencies initially equal. We also show the evolution of 
the mean number of coexisting alleles from starts with $M=3$ and $M=4$ 
alleles.  In both cases, the exact formul\ae\ correspond extremely well 
with the simulations, illustrating the utility of the diffusion approximation 
to the discrete population dynamics.

\medskip

\underline{Mean time to the $r$th extinction.} By solving the
appropriate backward Kolmogorov equation, one can straightforwardly
find the mean time to the first (and only) extinction event among two
alleles to be
\begin{equation}\label{tau_1}
\tau(x_0)=-2\left[ x_0\ln x_0 + (1-x_0) \ln (1-x_0) \right]\,.
\end{equation}
A combinatorial analysis, performed in \cite{lit75b} and which we will
summarise in Section~\ref{other2}, reveals the mean time to the $r^{\rm
th}$ extinction can be calculated from an $M=2$ result.  One finds
\begin{equation}\label{tau_r}
\tau_r(M)= -2 \sum_{s=r}^{M-1}(-1)^{s-r} \binom{s-1}{r-1}
\sum [(x_{i_1,0} + \ldots +x_{i_{k},0})\ln(x_{i_1,0} + \ldots
+x_{i_{s},0})] \;,
\end{equation}
where $x_{i_{k},0}$ is the initial fraction of allele $a_k$ present in the 
system. The second sum is over all possible $s$-subsets $\{ i_1, i_2,
\ldots, i_s \}$ of the $M$ alleles initially present in the population.

\medskip

\underline{Probability of a particular sequence of extinctions.}  Not
all extinction probabilities can be calculated by reduction to an
equivalent a two-allele problem.  A notable example is the probability
that allele $a_{i_1}$ goes extinct first, followed by $a_{i_2},
a_{i_3}, \ldots, a_{i_{M-1}}$ leaving only $a_{i_M}$ in the final
population, since in this case we ask for \emph{all}, rather than just
\emph{some} of a subset to be present in the population at a given
time.  This probability is given by
\begin{multline}\label{QP}
Q_{i_1,i_2,\ldots,i_{M-1}}(\vec{x}) = x_{i_M,0}
\frac{x_{i_{M-1},0}}{1-x_{{i_M},0}}
\frac{x_{i_{M-2},0}}{1-x_{{i_M},0}-x_{i_{M-1},0}} \cdots\\
\cdots\frac{x_{{i_2},0}}{1 - x_{{i_M},0} - x_{i_{{M-1},0}} - \cdots -
x_{{i_3},0}}
\end{multline}
in which $x_{i,0}$ is the initial frequency of allele $a_i$ in the
population.  This result appears in \cite{lit78a}, albeit without an
explicit derivation. We shall provide one in Section~\ref{other2}.

\medskip

\underline{First-passage probability.} An immediate consequence of the
previous result is the probability that a particular allele, $a_i$ is
the \emph{first} to become extinct, that is, at a time when all other
alleles have a nonzero frequency.  This is obtained by summing
(\ref{QP}) over all possible sequences of extinctions in which $a_i$
goes extinct first.  For example, when the initial number of alleles
$M=3$, the probability that $a_1$ goes extinct first can be found by
summing over the cases $(i_1,i_2,i_3) = (1,2,3), (1,3,2)$ in
(\ref{QP}).  One finds
\[
Q_1(x_{1,0},x_{2,0})=x_{2,0}\frac{x_{3,0}}{1-x_{2,0}}+x_{3,0}
\frac{x_{2,0}}{1-x_{3,0}}\,,
\]
where $x_{3,0}=1-x_{1,0}-x_{2,0}$.  This agrees with the result in
\cite{lit75b}.  By a similar method, one could determine the probability
for allele $a_i$ to be the \emph{second} to go extinct or any other
quantity of this type.

\subsubsection{Stationary distribution}

When mutations are absent, we see from (\ref{f_1}) that allele $a_i$
fixes with a probability equal to its initial frequency in the
population.  Hence in this case, the stationary distribution
\begin{equation}
P^\ast(\und{x}) = \lim_{t \to \infty} P(\und{x}, t |
\und{x}_0, 0)
\end{equation}
is zero everywhere except at those points that correspond to fixation
of a single allele.

On the other hand, when all the mutation rates $m_i$ at which all 
alleles $a_j, j \ne i$ mutate to $a_i$ are nonzero, the stationary
distribution is nonzero everywhere.  Furthermore, this distribution is
reached from any initial condition and takes the well-known form 
\cite{wri45,kim64b}
\begin{equation}
\label{Pstarmut}
P^\ast(\und{x}) = \Gamma(2R) \prod_{i=1}^M 
\frac{x_i^{2m_i-1}}{\Gamma(2m_i)} \;,
\end{equation}
where in keeping with the 2 allele case $R = R_1= \sum_{j=1}^{M} m_j$. 
Recall that the frequency $x_M$ is
implied through the normalisation $\sum_{i=1}^{M} x_i = 1$.

\subsubsection{Moments and Heterozygosity}

One way to calculate moments of the distribution is to integrate the
exact solution we have obtained above.  However, it is also possible
to derive differential equations for the moments in terms of
lower-order moments from the Kolmogorov equation, a procedure that
leads more quickly to simple, exact expressions for low-order
moments.  We quote some examples for general $M$ here, noting that
previously results for $M=2$ and $3$ without mutations have previously
appeared in \cite{kim55b,cro70} and for $M=2$ with mutations in
\cite{cro56,kim64b}.  It is also worth remarking that by calculating
moments, Kimura reconstructed the solution for the pdf for the case
$M=3$ and no mutations \cite{kim55b}.

\medskip

\underline{No mutations.} When mutations are absent, the mean of $x_i$
is conserved by the dynamics, i.e.,
\begin{equation}
\langle x_i(t) \rangle  = x_{i,0}\,.
\end{equation}
Meanwhile, the variance changes with time as
\begin{eqnarray}
\langle \langle x_i^{2} (t) \rangle\rangle_{x_i(0)=x_{i,0}} &\equiv& 
\langle x_i^{2} (t) \rangle_{x_i(0)=x_{i,0}} 
- \left( \langle x_i (t) \rangle_{x_i(0)=x_{i,0}} \right)^{2} \nonumber \\
&=& x_{i,0}(1-x_{i,0})\left( 1- e^{-t}\right) \;.\label{xvar_neut}
\end{eqnarray}
This result is clear, at least in the case $M=2$. The delta-function initial 
condition has zero variance, and the stationary distribution comprises
delta-functions at $x=0$ and $x=1$ with weights $(1-x_{0})$ and $x_0$
respectively.  For the latter one then has $\langle x^{2} \rangle =
(1-x_{0}) \cdot 0 + x_{0} \cdot 1 = x_{0}$ and so $\langle \langle
x^{2} \rangle \rangle = x_{0}(1 - x_{0})$.

The covariance, which does not appear in the two allele system is:
\begin{equation}
\langle \langle x_ix_j \rangle\rangle \equiv \langle x_ix_j \rangle-
\langle x_i (t) \rangle\langle x_j (t) \rangle \nonumber =
x_{i,0}x_{j,0}(e^{-t}-1)\,.
\end{equation}

\medskip

\underline{With mutations.} Crow and Kimura \cite{cro56} give moments
only for the biallelic case, and then in terms of hypergeometric
functions. We provide simpler results in explicit form here, again for
any $M$.  The mean frequency of any allele behaves as
\begin{equation}
\label{meanx_M}
\langle x_i(t) \rangle = \frac{m_i}{R} + \left( x_{i,0} - \frac{m_i}{R}
\right) e^{-R t} = \eta_{i} + \zeta_{i} e^{-R t}\;,
\end{equation}
where
\begin{equation}
\eta_{i} = \frac{m_i}{R}, \ \ \ \zeta_{i} = x_{i,0} - \frac{m_i}{R}\,.
\label{eta_zeta}
\end{equation}
The mean (\ref{meanx_M}) exponentially approaches the fixed point of the 
deterministic motion. Recalling that $R=\sum_im_i$, we see in 
Eq.~(\ref{Kol_2_mut}) the function $A_i(x)$ vanishes at the point 
$x=m_i/R \;\forall i$.

The variance is more complicated:
\begin{multline}
\langle\langle x_i^2(t)\rangle\rangle =
\frac{\eta_{i}(1-\eta_{i})}{(2R+1)} + \frac{\zeta_{i}(1-2\eta_{i})}{(R+1)} 
e^{-Rt} - \zeta_{i}^{2} e^{-2Rt} \\
- \left\{ \frac{\eta_{i}(1-\eta_{i})}{(2R+1)} + 
\frac{\zeta_{i}(1-2\eta_{i})}{(R+1)} - \zeta_{i}^{2} \right\} e^{-(2R+1)t}\,.
\label{varx_M}
\end{multline}
As $t\to\infty$, this converges to the limit
\begin{equation}
\langle\langle
x^2(t)\rangle\rangle_{t\to\infty}= \frac{\eta_{i}(1-\eta_{i})}{(2R+1)} =
\frac{m_i(R-m_i)}{R^2(2R+1)} \;.
\end{equation}

\begin{figure}[tb]
\begin{center}
\includegraphics[width=0.45\linewidth]{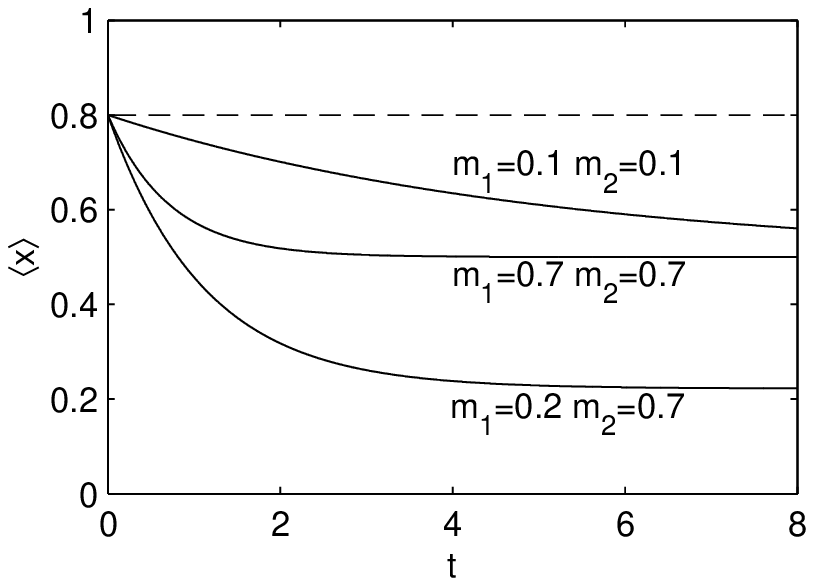}
\hspace{0.05\linewidth}
\includegraphics[width=0.45\linewidth]{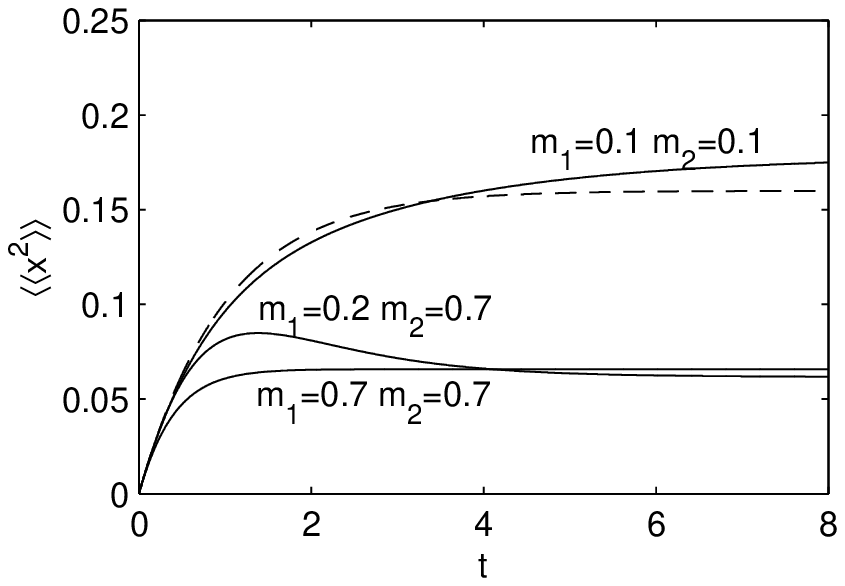}
\end{center}
\caption{\label{mean_var} Solid lines: Time evolution of the mean
(left) and variance (right) for a biallelic system for three different
mutation parameter combinations, and with $x_0=0.8$. Dashed lines:
Evolution of mean and variance when mutation is absent. }
\end{figure}

In Fig.~\ref{mean_var} we show the mean and variance as a function of
time in the biallelic case, for three different pairs of mutation
parameters. The mean approaches the stationary distribution value
exponentially, at a faster rate the larger the values of the mutation
parameters. When $m_1=m_2$, the variance rises asymptotically to the
stationary limit, again faster when the mutation parameters are
larger. Notice that for small values of $m_1$ and $m_2$, the final
variance is close to that when no mutations occur, while the larger the
mutation parameters become, the narrower the final distribution. When
the parameters are unequal, the variance can reach a maximum value
before descending slowly to the final limit.

\begin{figure}[tb]
\begin{center}
\includegraphics[width=0.55\linewidth]{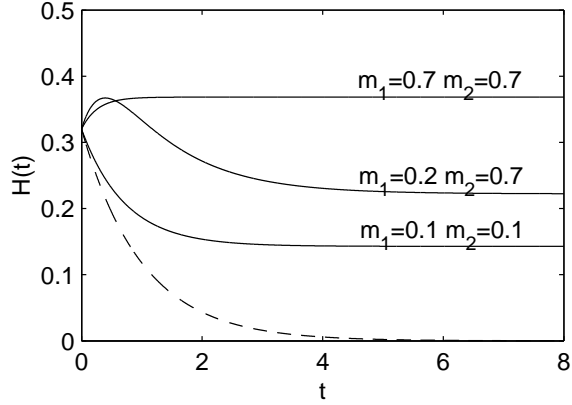}
\end{center}
\caption{\label{H_t} The time development of the heterozygosity with
two alleles, for the case with no mutations (dashed curve) and for three
different cases of mutation (solid curves), $x_0=0.8$ in each case. }
\end{figure}

The covariance (with $i \neq j$) is
\begin{eqnarray}
\langle\langle x_ix_j\rangle\rangle &\equiv&
\langle x_ix_j\rangle-\langle x_i\rangle\langle x_j\rangle \nonumber\\
&=& - \frac{\eta_{i} \eta_{j}}{(2R+1)} - \frac{(\eta_{i}\zeta_{j}
+\eta_{j}\zeta_{i})}{(R+1)} e^{-Rt} - \zeta_{i} \zeta_{j} e^{-2Rt} \nonumber\\
&& {} + \left\{ \frac{\eta_{i} \eta_{j}}{(2R+1)} + 
\frac{(\eta_{i}\zeta_{j}+\eta_{j}\zeta_{i})}{(R+1)} + 
\zeta_{i} \zeta_{j} \right\} e^{-(2R+1)t}\,.
\label{covarx_M}
\end{eqnarray}

\medskip

\underline{Heterozygosity.}
As asserted by Kimura \cite{kim55b}, in the absence of mutation the
heterozygosity decays as $e^{-t}$ regardless of the initial number of
alleles.  The heterozygosity is the probability that two alleles
chosen at random from the population will be different. When there are
multiple alleles, this probability can be calculated from the first
and second moments using
\begin{equation}\label{H_sum}
H(t)=\sum_i[2(\langle x_i\rangle-\langle x_i^2\rangle)-
\sum_{j\neq i}\langle x_ix_j\rangle]\;.
\end{equation}
When expressions for the various moments are substituted into this expression,
it is found that, in the absence of mutations, the constant terms cancel, 
leaving only terms decaying as $e^{-t}$. Therefore we have
\begin{equation}\label{H_neut}
H(x_{0},t)=H_0e^{-t}\;,
\end{equation}
where we find $H_0=2\sum_i(x_{i,0}-x_{i,0}^2)-\sum_i\sum_{j\neq
i}x_{i,0}x_{j,0}$. Kimura derived this result from the expected
change in $H$ with each generation.

The calculation with mutation present, however, does not give such a
simple result. The heterozygosity can also be found for any number of
alleles when mutation is present. However, the form becomes
increasingly complicated as $M$ increases. Therefore we give as an
example only the two allele result here:
\begin{eqnarray}
H(x_{0},t) &\equiv& 2\langle x(1-x)\rangle \nonumber \\
&=& \frac{4m_1m_2}{R(2R+1)} - 2\left(\frac{m_2-m_1}{R+1}\right)\left(x_{0}-
\frac{m_1}{R}\right) e^{-Rt} \nonumber \\
&& {} - 2\left[ x_{0}^2 - \left(\frac{2m_1+1}{R+1} \right) 
\left(x_{0} - \frac{m_1}{2R+1} \right) \right] e^{-(2R+1)t} \,.
\label{H_mut}
\end{eqnarray}
This is plotted in Fig.~\ref{H_t}. We see that the presence of
mutation between the two alleles maintains the heterozygosity at a
finite level.

\section{Solution of the Kolmogorov equation}
\label{soln}

In this section we will derive the solution of the Kolmogorov equation
(\ref{Kol_M_u}), which was given in Section \ref{cond_pdf}.  In
particular we will derive the differential equation that is satisfied
by every single one of the factors $\psi^{(i)}(u_i)$ that appear in
the eigenfunction $\Phi_{\lambda}(\vec{u})$ in
Eq.~(\ref{soln_M_second}) which applies when no mutations occur.  We
shall also present the more general differential equation that must be
solved to determine the corresponding factors in
(\ref{soln_M_second_m}) that give the eigenfunctions when mutations
are present.  The more technical points of the discussion are
relegated to Appendices B and C, in order to avoid interrupting the
flow of the arguments.

\subsection{Change of variables}

It is worthwhile to recall briefly the derivation of the Kolmogorov
equation itself, since it will turn out that the most efficient way of
changing from the $\und{x}$ variables to the $\und{u}$ variables is go
back to an early stage of this derivation. The starting point for the
derivation is the Chapman-Kolmogorov equation which is one of the
defining equations for Markov processes:
\begin{equation} 
P(\und{x}, t'|\und{x}_{0}, 0) = \int \D\und{x}' P(\und{x}, t' |\und{x}', t)\,
P(\und{x}', t|\und{x}_{0}, 0)\,.
\label{C-K_eqn}
\end{equation}
If we take $t' = t+\Delta t$ and expand 
\begin{equation} 
P(\und{x}, t+\Delta t |\und{x}', t)\,
\label{expand}
\end{equation}
in powers of $\Delta x_{i} (t)=x_{i} (t+\Delta t) - x_{i} (t)$, we 
find---in the limit $\Delta t \to 0$---the Kramers-Moyal expansion for 
$P(\und{x}, t|\und{x}_{0}, 0)$~\cite{ris89}:
\begin{equation}
\frac{\partial P}{\partial t} = \sum^{\infty}_{m=1}\, \frac{(-1)^{m}}{m!}
\frac{\partial^m}{\partial x_{i_1} \ldots \partial x_{i_m}} 
\left\{ \alpha_{i_{1}\ldots i_{m}} (\und{x}) P \right\}\,.
\label{KM_expansion}
\end{equation}
Here the $\alpha_{i_{1}\ldots i_{m}} (\und{x})$ are the jump moments defined 
through the relation
\begin{equation}
\left\langle \Delta x_{i_{1}} (t) \Delta x_{i_{2}} (t) \ldots 
\Delta x_{i_{m}} (t) \right\rangle_{\und{x}(t)=\und{x}} = 
\alpha_{i_{1}\ldots i_{m}} (\und{x})\Delta t + \mathcal{O}(\Delta t)^{2}\,.
\label{jm_defn}
\end{equation}
The angle brackets are averages over realisations of the process $\und{x} (t)$;
$\und{x} (t)$ is a stochastic process, whereas $\und{x}_0$ and $\und{x}$ 
are simply the initial and final states. 

So far very little has been assumed other than the Markov property of
the stochastic process. The Markov nature of the process follows from
the fact that the sampling of gametes from the gamete pool depends
only on its current state, and not on its past history. If we assume
that no deterministic forces are present, only the random mating
process, then $\langle \Delta x_{i_1} (t) \rangle_{\und{x}(t)=\und{x}}
= 0$. If deterministic forces are present, then there will be an
$\mathcal{O}(\Delta t)$ contribution to $\langle \Delta x_{i_1} (t)
\rangle_{\und{x}(t)=\und{x}}$ which is equal to
$(\D x_{i_1}/\D t)|_{\rm det}\, \Delta t$. Here $(\D x_{i_1}/\D t)|_{\rm det}$
is the rate of change of $x_{i_1}$ in the deterministic ($N \to
\infty$) limit.  The higher jump moments reflect the statistics of the
sampling process. In the case of two alleles this is binomial and so
the second jump moment is proportional to $x(1-x)$, which is the
variance of this distribution.  Within the diffusion approximation, a
unit of time is taken to be $2N$ generations, so that the time between
generations is $\Delta t = 1/(2N)$. This implies that all jump moments
higher than the second are higher order in $\Delta t$. This gives the
results (\ref{Kol_2_mut}) and (\ref{AandD_2}). For the case of general
$M$, the multinomial distribution applies, which again means that the
$\alpha_{i_{1}\ldots i_{m}} (\und{x})$ for $m > 2$ vanish, and the
second order jump moment is given by (\ref{diff_M}), up to a factor of
$2$.

This is the briefest summary of the derivation of Eq.~(\ref{Kol_M_mut}). We 
refer the reader to \cite{ris89} for a fuller account. However, the above 
discussion is sufficient for our purposes, which is to note that the derivation
involving Eqs.~(\ref{C-K_eqn})-(\ref{jm_defn}) may be carried out for any 
coordinate system, and specifically in the coordinate system
$\und{u}=(u_{1}, u_{2},\ldots, u_{M-1})$ for which the Kolmogorov equation is 
separable. In this case the Kramers-Moyal expansion for 
$\mathcal{P} (\und{u}, t|\und{u}_{0}, 0)$ is
\begin{equation}
\frac{\partial \mathcal{P}}{\partial t} = \sum^{\infty}_{m=1}\, 
\frac{(-1)^{m}}{m!}\frac{\partial^m}{\partial u_{i_1} \ldots \partial u_{i_m}} 
\left\{ \tilde{\alpha}_{i_{1}\ldots i_{m}} (\und{u}) \mathcal{P} \right\}\,,
\label{KM_expansion_new}
\end{equation}
where $\tilde{\alpha}_{i_{1}\ldots i_{m}} (\und{u})$ are the jump moments 
in the new coordinate system:
\begin{equation}
\left\langle \Delta u_{i_{1}} (t) \Delta u_{i_{2}} (t) \ldots 
\Delta u_{i_{m}} (t) \right\rangle_{\und{u}(t)=\und{u}} = 
\tilde{\alpha}_{i_{1}\ldots i_{m}} (\und{u})\Delta t + 
\mathcal{O}(\Delta t)^{2}\,.
\label{jm_defn_new}
\end{equation}
The most straightforward way of deriving the Kolmogorov equation in the 
$\und{u}$ variables is to start from Eq.~(\ref{KM_expansion_new}) and determine
the $\tilde{\alpha}_{i_{1}\ldots i_{m}} (\und{u})$ by relating them to the 
known $\alpha_{i_{1}\ldots i_{m}} (\und{x})$. This is carried out in Appendix 
B and the result is the equations (\ref{Kol_M_u}) and (\ref{calA_calD}).

\subsection{Separation of variables}

The next step in the solution of the Kolmogorov equation is to
separate out the time and $u_i$ coordinates.  That is, we write the
general solution as a linear combination of such solutions:
\begin{equation}
\mathcal{P} (\und{u}, t|\und{u}_{0}, 0) = \sum_{\lambda} a_{\lambda} 
(\und{u}_{0}) \Phi_{\lambda} (\und{u}) e^{-\lambda t}\,,
\label{lin_comb}
\end{equation}
where the function $\Phi_{\lambda} (\und{u})$ satisfies
\begin{equation}
\sum_{i} \frac{1}{\prod_{j<i} (1-u_j)}\, \frac{\partial }{\partial u_i}
\left\{ \left( R_{i} u_{i} - m_{i} \right) + \frac{1}{2} \frac{\partial } 
{\partial u_i} \left[ u_{i}(1-u_i) \right] \right\} 
\Phi_{\lambda} (\und{u}) = - \lambda \Phi_{\lambda} (\und{u})
\label{eigen_eqn}
\end{equation}
and $R_i = \sum_{j=i}^{M} m_j$.  We may apply initial conditions directly 
on (\ref{lin_comb}):
\begin{equation}
\delta (\und{u}-\und{u}_{0}) = \sum_{\lambda} a_{\lambda} (\und{u}_{0})
\Phi_{\lambda} (\und{u})\,.
\label{initial_conds}
\end{equation}
By assuming the orthogonality relation
\begin{equation}
\int \D\und{u}\,w(\und{u}) \Phi_{\lambda} (\und{u}) \Phi_{\lambda '} (\und{u})
= \delta_{\lambda \lambda'}\,,
\label{ortho_1}
\end{equation}
where $w(\und{u})$ is an appropriate weight function, we then have
that $a_{\lambda} (\und{u})=w(\und{u}) \Phi_{\lambda} (\und{u})$ and
so
\begin{equation}
\mathcal{P} (\und{u}, t|\und{u}_{0}, 0) = w(\und{u}_{0}) \sum_{\lambda} 
\Phi_{\lambda} (\und{u}_{0}) \Phi_{\lambda} (\und{u}) e^{-\lambda t}\,.
\label{post_initial_conds}
\end{equation}

To determine $\Phi_{\lambda} (\und{u})$ we have in principle to solve
the partial differential equation (\ref{eigen_eqn}) in $n=M-1$
variables. We will now prove that this equation is separable in the
$\und{u}$ variables.  For clarity we will start where mutation does
not take place (that is, omit the factor $( R_{i} u_{i} - m_{i})$ in
(\ref{eigen_eqn}))---the argument goes through in an identical
fashion when this term is present, as we will discuss later.

When $n=1\ (M=2)$ $\Phi_{\lambda} (\und{u})$ satisfies the equation 
\begin{equation}
\frac{1}{2} \frac{\D^2}{\D u^2_1} \left[ u_{1}(1-u_1) 
\Phi_{\lambda^{(1)}} (u_1) \right] = - \lambda^{(1)} 
\Phi_{\lambda^{(1)}} (u_1)\,,
\label{neutral_1}
\end{equation}
where we have introduced the superscript $(1)$ on to the eigenvalue $\lambda$ 
to identify it as belonging to the $n=1$ problem. This ordinary differential 
equation may be solved in a straightforward fashion as we discuss in
the next subsection below.

When $n=2\ (M=3)$ $\Phi_{\lambda} (\und{u})$ satisfies the equation 
\begin{eqnarray}
\frac{1}{2} \frac{\partial }{\partial u^2_1} \left[ u_{1}(1-u_1)
\Phi_{\lambda^{(2)}} (u_1, u_2) \right] &+& \nonumber \\
\frac{1}{2}\,\frac{1}{1-u_1}\,\frac{\partial }{\partial u^2_2} 
\left[ u_{2}(1-u_2) \Phi_{\lambda^{(2)}} (u_1, u_2) \right] &=& 
 - \lambda^{(2)} \Phi_{\lambda^{(2)}} (u_1, u_2)\,.
\label{neutral_2}
\end{eqnarray}
We look for a separable solution of the form
$\Phi_{\lambda^{(2)}} (u_1, u_2)= \psi (u_1) \phi_{\lambda^{(1)}} (u_2)$ where
\begin{displaymath}
\frac{1}{2} \frac{\D^2}{\D u^2_2} \left[ u_{2}(1-u_2) 
\phi_{\lambda^{(1)}} (u_2) \right] = - \lambda^{(1)} 
\phi_{\lambda^{(1)}} (u_2)\,,
\end{displaymath}
that is, $\phi_{\lambda^{(1)}} (u_1)$ satisfies the equation corresponding to
the $n=1$ problem. It is straightforward to show that a separable solution
exists if $\psi (u_1)$ satisfies the equation
\begin{equation}
\frac{1}{2} \frac{\D^2}{\D u^2_1} \left[ u_{1}(1-u_1) 
\psi_{\lambda^{(2)};\lambda^{(1)}} (u_1) \right] = \left\{ - \lambda^{(2)} 
+ \frac{\lambda^{(1)}}{(1-u_1)} \right\} \psi_{\lambda^{(2)};\lambda^{(1)}} 
(u_1)\,.
\label{psi_eqn_1}
\end{equation}
We have introduced the subscript $\lambda^{(2)};\lambda^{(1)}$ on $\psi (u_1)$
to show that it depends on both the eigenvalues of the $n=2$ and the $n=1$ 
problems. As we will see, it is a remarkable fact that the solution of the
problem with $M=n+1$ alleles only involves the function $\psi$ appearing in
Eq.~(\ref{psi_eqn_1}). Since $\phi_{\lambda^{(1)}} (u_2)$ may also be
written as $\psi_{\lambda^{(1)};0}$ we may write the separable form of the 
solution in the $n=2$ case as 
\begin{equation}
\Phi_{\lambda^{(2)}} (u_1, u_2)= \psi_{\lambda^{(2)};\lambda^{(1)}} (u_1) 
\psi_{\lambda^{(1)};0} (u_2)\,.
\label{sep_soln_2}
\end{equation} 
We can now prove that the general case of $M=n+1$ alleles is separable as 
follows. We look for a solution of (\ref{eigen_eqn}) of the form
\begin{displaymath}
\Phi_{\lambda^{(n)}} (u_1, u_2,\ldots,u_n)= \psi (u_1) 
\Phi_{\lambda^{(n-1)}} (u_2,\ldots,u_{n})\,, 
\end{displaymath}
where $\Phi_{\lambda^{(n-1)}} (u_2,\ldots,u_{n})$ is the solution of 
(\ref{eigen_eqn}) with $n-1$ alleles, but with the replacements 
$u_{1} \to u_{2}, u_{2} \to u_{3},\ldots,u_{n-1} \to u_{n}$. By explicit
substitution into (\ref{eigen_eqn}) it is not difficult to demonstrate this,
and to find that $\psi (u_1)$ satisfies Eq.~(\ref{psi_eqn_1}), where now 
$\lambda^{(2)}$ and $\lambda^{(1)}$ are replaced by $\lambda^{(n)}$ and 
$\lambda^{(n-1)}$ respectively. Therefore we have shown that
\begin{equation}
\Phi_{\lambda^{(n)}} (u_1, u_2,\ldots,u_n)= 
\psi_{\lambda^{(n)};\lambda^{(n-1)}} (u_1) 
\Phi_{\lambda^{(n-1)}} (u_2,\ldots,u_{n})\,.
\label{sep_soln} 
\end{equation}
This shows that if the $n-1$ allele problem is separable, then the $n$ allele
problem is separable. But Eq.~(\ref{sep_soln_2}) shows that the $n=2$ problem 
is separable, and so by induction the problem for general $n$ is separable. The
explicit solution is
\begin{equation}
\Phi_{\lambda^{(n)}} (u_1, u_2,\ldots,u_n)= \prod^{n}_{i=1} 
\psi_{\lambda^{(n+1-i)};\lambda^{(n-i)}} (u_i)\,,
\label{sep_soln_n} 
\end{equation}
where $\lambda^{(0)} \equiv 0$ and where $\psi_{\lambda;\lambda'} (u)$
satisfies the equation
\begin{equation}
\frac{1}{2} \frac{\D^2}{\D u^2} \left[ u (1-u) 
\psi_{\lambda;\lambda'} (u) \right] = \left\{ - \lambda 
+ \frac{\lambda'}{(1-u)} \right\} \psi_{\lambda;\lambda'} (u)\,.
\label{psi_eqn}
\end{equation}
Thus we have finally shown that in order to obtain the eigenfunction
$\Psi_{\lambda^{(n)}}(u_1, \ldots, u_n)$ it is necessary only to solve
this \emph{single} ordinary differential equation for general
$\lambda$ and $\lambda^\prime$.  The full eigenfunction is then just a
product of solutions of this equation.

A similar equation arises from an analogous argument when mutation is
included, that is, the factor $( R_{i} u_{i} - m_{i})$ is restored
into Eq.~(\ref{eigen_eqn}): all that is required is to make the
substitution
\begin{displaymath}
\frac{1}{2} \frac{\partial }{\partial u_i} \left[ u_{i}(1-u_i) \right] 
\longrightarrow \frac{1}{2} \frac{\partial } {\partial u_i} 
\left[ u_{i}(1-u_i) \right] + \left( R_{i} u_{i} - m_{i} \right)
\end{displaymath}
in the above argument. The only difference is that now the function $\psi$
explicitly depends on $i$ because of the existence of the terms $R_i$ and 
$m_i$. Thus the results corresponding to Eqs.~(\ref{sep_soln_n}) and
(\ref{psi_eqn}) are:
 \begin{equation}
\Phi_{\lambda^{(n)}} (u_1, u_2,\ldots,u_n)= \prod^{n}_{i=1} 
\psi^{(i)}_{\lambda^{(n+1-i)};\lambda^{(n-i)}} (u_i)\,,
\label{sep_soln_n_mut} 
\end{equation}
where $\psi^{(i)}_{\lambda;\lambda'} (u_i)$ satisfies the equation
\begin{eqnarray}
\frac{\D}{\D u_i} \left[ \left( R_{i} u_{i} - m_{i} \right) 
\psi^{(i)}_{\lambda;\lambda'} (u_i) \right] &+& 
\frac{1}{2} \frac{\D^2}{\D u^2_i} \left[ u_i(1-u_i) 
\psi^{(i)}_{\lambda;\lambda'} (u_i) \right] \nonumber \\
&=& \left\{ - \lambda + \frac{\lambda'}{(1-u_i)} \right\} 
\psi^{(i)}_{\lambda;\lambda'} (u_i)\,.
\label{psi_eqn_mut}
\end{eqnarray}
The rest of this section will be devoted to solving the ordinary
differential equations (\ref{psi_eqn}) and (\ref{psi_eqn_mut}) subject
to the boundary conditions of the problem.

\subsection{Solution of the ordinary differential equations}

\subsubsection{Reduction to a standard form}

To solve the ordinary differential equation (\ref{psi_eqn_mut}) (which
includes the less general Eq.~(\ref{psi_eqn}) as a special case) it is
useful to cast it in a standard form.  To this end, we introduce the
function $f^{(i)}_{\lambda;\lambda'} (u_i)$ by writing
\begin{equation}
\psi^{(i)}_{\lambda;\lambda'} (u_i) = \left( 1 - u_{i} \right)^{\gamma_i}\,
f^{(i)}_{\lambda;\lambda'} (u_i)\,,
\label{f_defn}
\end{equation}
where $\gamma_i$ is a constant which is to be appropriately chosen shortly. 
Substituting Eq.~(\ref{f_defn}) into Eq.~(\ref{psi_eqn_mut}) one finds
(dropping the subscripts and superscripts on $f$)
\begin{multline}
u_{i}(1-u_i) f'' + 2\left[ (1-m_i) + (R_{i} - 
(\gamma_{i}+2)u_{i} \right] f' \\ +
\left[ 2\lambda - (\gamma_{i}+1)(\gamma_{i}+2) + 
2R_{i}(\gamma_{i}+1) \right] f  \\
= \frac{\left[2\lambda' - \gamma_{i}(\gamma_{i} +1) - 
2\gamma_{i}(m_{i}-R_{i})\right]}{1-u_{i}}\,f\,. 
\label{f_eqn}
\end{multline}
We will choose $\gamma_i$ so that the term involving $(1-u_i)$ in the 
denominator vanishes, that is, we ask that
\begin{equation}
\gamma_{i}\left( \gamma_{i}+1+2m_{i}-2R_{i}\right) = 2\lambda'\,.
\label{gammai_choice}
\end{equation}
This is a quadratic equation for $\gamma_i$, but it is simple to see that if
$\gamma_i$ is one solution, the second one is simply 
$\gamma_{i}' = 2R_{i}-2m_{i}-1-\gamma_{i}$. It will turn out that either choice
leads to the same form for $\psi^{(i)}_{\lambda;\lambda'} (u_i)$.

If we choose $\gamma_i$ to satisfy Eq.~(\ref{gammai_choice}), then 
Eq.~(\ref{f_eqn}) has the form
\begin{equation}
u_i(1-u_i)f'' + \left[ c - (a+b+1)u_i \right] f' - ab f = 0\,,
\label{hypergeo}
\end{equation}
where
\begin{equation}
c = 2(1-m_i);\ \ a+b=2\gamma_{i}+3-2R_{i};\ \ ab=(\gamma_{i}+1)(\gamma_{i}+2)
-2R_{i}(\gamma_{i}+1) - 2\lambda\,.
\label{abc_defn}
\end{equation}
The solution of the equation has now been reduced to a standard form,
since Eq.~(\ref{hypergeo}) is the hypergeometric
equation~\cite{abr65}, whose solutions are hypergeometric functions
$F(a,b,c,z)$.  The details of the analysis of this equation are
given in Appendix C; here we only describe the general features and
specific form of the solution, without going into their
derivation. That is, we find the constants $\gamma_i$ and
$\lambda$. In the following discussion it is convenient to treat the 
cases where mutations do and do not occur separately.

\subsubsection{Solution in the absence of mutations}
\label{soln_neutral}
We begin with the situation where mutations are absent, and so the pdf
is found by solving Eq.~(\ref{hypergeo}), but where now
\begin{equation}
c = 2;\ \ a+b=2\gamma_{i}+3;\ \ ab=(\gamma_{i}+1)(\gamma_{i}+2) - 2\lambda\,.
\label{abc_defn_neut}
\end{equation}
The $\gamma_i$ are chosen so that $\gamma_{i}(\gamma_{i}+1) =
2\lambda'$. We begin the discussion, for orientation, by assuming that
only two alleles are present. This amounts to solving
Eq.~(\ref{neutral_1}) or Eq.~(\ref{psi_eqn}) with $\lambda' = 0$, and
the solution is therefore the function $\psi_{\lambda^{(1)};0}
(u_1)=\phi_{\lambda^{(1)}} (u_1)$. This function also appears in the
solution for arbitrary $M$, given by Eq.~(\ref{sep_soln_n}), and so
this result is also required in the general case.

Since in this simple case of two alleles $\lambda'=0$, it follows that
$\gamma_{1}=0$ or $\gamma_{1}=-1$. As discussed in Appendix C, either
choice gives the same solution, and we therefore take
$\gamma_{1}=0$. We have now only to solve the second order ordinary
differential equation (\ref{hypergeo}) subject to boundary conditions
at $u_{1}=0$ and $u_{1}=1$. As is familiar in such problems \cite{mor53}, 
the general solution is first expressed as a linear
combination of two independent solutions. One of these is ruled out by
one of the boundary conditions (e.g. the one at $u_{1}=0$) and the
application of the other boundary condition constrains a function of
$\lambda$ to be an integer; in our case one finds $a=-l,\
l=0,1,\ldots$. Details are in Appendix C, but we do wish to mention
the nature of the boundary conditions here.  Technically \cite{fel52}
these are \emph{exit} boundary conditions which means that all the
probability which diffuses to the boundary is ``extracted'' from the
interval $(0,1)$.  Na\"{\i}vely, one might imagine that it is appropriate
to impose absorbing boundary conditions, i.e., that the pdf vanishes
at $u_1=0$ and $u_1=1$.  However, since the diffusion coefficient,
defined in Eq.~(\ref{Kol_2}), vanishes on the boundaries the diffusion
equation itself imposes some sort of absorption.  A more careful
mathematical analysis, such as the informal argument presented in
Appendix A, reveals that the appropriate constraint on the eigenfunctions 
is that they should diverge less rapidly than a square-root at both 
boundaries.

The result $a=-l$ imposed by the boundary conditions together with the
definitions (\ref{abc_defn_neut}) imply that $b=l+3, ab= -l^{2}-3l$
and so
\begin{equation}
\lambda = \frac{1}{2} \left( l+1 \right) \left( l+2 \right)\,.
\label{lambda_1_quant}
\end{equation}
The solutions of the differential equation will be labelled by the integer
$l$ and consist of polynomials of order $l$. These have already been mentioned 
in Section \ref{overview} (Eq.~(\ref{soln_2})), whilst more technical
information can be found in Appendix~C.

Now we can move on to the solution for general $M$. As explained above it is
sufficient to solve Eq.~(\ref{psi_eqn}) where $\lambda'$ is given by the 
$\lambda$ found in the solution of the $M-1$ allele problem. We have seen 
that for $M=2$, this is given by (\ref{lambda_1_quant}), and so this will
be the $\lambda'$ for the $M=3$ problem. We will see that this will have the
general structure $(L+1)(L+2)/2$, where $L$ is an integer, and therefore we 
take this form for $\lambda'$. From the way that the $M-1$ problem is embedded 
in the $M$ allele problem, we always have only to solve the $u_1$ equation
and so we need to determine $\gamma_1$. This is determined by the equation 
$\gamma_{1}(\gamma_{1}+1) = 2\lambda'$ from which we see that 
$\gamma_{1}=L+1$ or $\gamma_{1}=-L-2$. Again, both give the same result and 
so we take the former. Applying the boundary conditions again gives $a=-l$, 
with $l=0,1,\ldots$---see Appendix~C for the details. Using 
Eq.~(\ref{abc_defn_neut}) one sees that $b=2L+l+5$ and so
\begin{equation}
\lambda = \frac{1}{2} \left( L+l+2 \right) \left( L+l+3 \right) =
\frac{1}{2} \left( L' + 1 \right) \left( L' + 2 \right)\,,
\label{lambda_gen_quant}
\end{equation}
where $L' = L+l+1$. This justifies the choice for $\lambda'$. 

In summary, we have found that $\lambda^{(1)}=(l_{1}+1)(l_{1}+2)/2$, 
$\lambda^{(2)}=(l_{1}+l_{2}+2)(l_{1}+l_{2}+3)/2, \ldots$. If we define
\begin{equation}
L_{i} \equiv \sum_{j \leq i} \left( l_{j} + 1 \right)\,,
\label{capL_i}
\end{equation}
then $\lambda^{(i)} = L_{i}(L_{i}+1)/2$, and we have recovered
(\ref{lambda_and_L}). The general solution is given by
Eqs.~(\ref{soln_M_first})-(\ref{soln_M_fourth}) and discussed further
in Appendix C.

\subsubsection{Solution with mutations}
\label{soln_with_muts}

When mutations are present we have to solve Eq.~(\ref{hypergeo})
subject to the general set of parameter values given by
Eq.~(\ref{abc_defn}). While it is true that this slightly complicates
the analysis as compared with that of the mutation-free case, the most
significant change is the nature of the boundary conditions. The
introduction of mutations in the way we have done here renders the
boundaries reflecting, which are defined as having no probability
current through them. The current $J_{i} (\und{u}, t)$ satisfies the
continuity equation~\cite{ris89,gar04}
\begin{equation}
\frac{\partial \mathcal{P}}{\partial t} + \sum^{M-1}_{i=1} 
\frac{\partial J_{i}}{\partial u_{i}} = 0\,,
\label{continuity}
\end{equation}
where (compare with Eqs.~(\ref{Kol_M_u}) and (\ref{calA_calD})), 
\begin{eqnarray}
J_{i} (\und{u}, t) &=& \left[ \mathcal{A}_{i} (\und{u}) 
\mathcal{P} (\und{u},t) \right] - \frac{\partial }{\partial u_{i}} 
\left[ \mathcal{D}_{i}(\und{u}) \mathcal{P} (\und{u},t) \right] \nonumber \\
&=& \frac{1}{ \prod_{j<i}(1-u_j)} \left\{ \left( m_{i} - R_{i} u_{i} \right)
\mathcal{P} - \frac{1}{2} \frac{\partial }{\partial u_i} \left[ u_i(1-u_i) 
\mathcal{P} \right] \right\}\,.
\label{current_defn}
\end{eqnarray}
Using the separable form of the solution (\ref{sep_soln_n_mut}), and asking 
that the current is zero on the boundaries lead to the conditions
\begin{equation}
\left\{ 2\left( R_{i} u_{i} - m_{i} \right) 
\psi^{(i)}_{\lambda_{i};\lambda_{i-1}} (u_i) + \frac{\D}{\D u_i} 
\left[ u_i(1-u_i) 
\psi^{(i)}_{\lambda_{i};\lambda_{i-1}} (u_i) \right] \right\}_{u_{i}=0,1} 
= 0\,,
\label{reflectbc_psi}
\end{equation}
or in terms of the solutions of the hypergeometric equation
\begin{eqnarray}
\left\{ \left[ 2 \left( R_{i} u_{i} - m_{i} \right) + \left\{ 1 - 
(\gamma_{i}+2)u_{i} \right\} \right] \left( 1 - u_{i} \right)^{\gamma_i}   
f^{(i)}_{\lambda_{i};\lambda_{i-1}} (u_i) \right. &+& \nonumber \\  
\left. u_i \left( 1-u_i \right)^{\gamma_{i}+1} 
\frac{\D}{\D u_i} 
f^{(i)}_{\lambda_{i};\lambda_{i-1}}\right\}_{u_{i}=0,1} &=& 0\,.
\label{reflectbc_f}
\end{eqnarray}
As before, we will discuss the general aspect of the solution here,
deferring technical details to Appendix~C.  We begin with the case of
two alleles.

The solution to the Kolmogorov equation when only two alleles are
present has been given by Crow and Kimura~\cite{cro56,cro70} and in
our case corresponds to the solution of Eq.~(\ref{psi_eqn_mut}) with
$\lambda' =0$.  It then follows from Eq.~(\ref{gammai_choice}) that we
may take $\gamma_{1}=0$ (note that $i$ takes on the value $1$ in this
case, since there is only one variable in the problem: $u_{1}=x$). It
therefore follows from Eq.~(\ref{abc_defn}) that $c = 2(1-m_1),\
a+b=3-2R_{1}$ and $ab= 2(1-R_{1}-\lambda)$. We begin by examining the
nature of the solutions of Eq.~(\ref{hypergeo}) in the vicinity of
$u_{1}=0$. Three separate cases have to be examined: (i) $c$ not an
integer, (ii) $c=1$, and (iii) $c=0,-1,-2,\ldots$. When the boundary
condition (\ref{reflectbc_f}) is applied, one of the two independent
solutions is ruled out. The application of the other boundary
condition again imposes a condition on $a$. In Appendix~C, we show
that
\begin{equation}
\lambda = \frac{1}{2} l \left( 2R_{1} + l -1 \right)\,,
\label{lambda_1_quant_mut}
\end{equation}
where $l$ is a non-negative integer and, as $M=2$, $R_{1} =
m_{1}+m_{2}$. The solutions of the differential equations are again
labelled by an integer $l$, and once again turn out to be Jacobi
Polynomials which have been given in Section \ref{overview}.

When determining the solution for general $M$, the eigenvalue
$\lambda'$ in Eq.~(\ref{psi_eqn_mut}) is given by the eigenvalue
$\lambda$ found in the $M-1$ allele problem, as occurred in the special
case of no mutations. However, more care is required here, since the
iterative nature of the problem manifests itself in such a way that
the solution with $M$ alleles is determined in terms of a function
$\psi (u_1)$ and the solution with $M-1$ alleles but with $u_{i}
\rightarrow u_{i+1}, m_{i} \rightarrow m_{i+1}$ and $R_{i} \rightarrow
R_{i+1}$, where $i=1,2\ldots,M-1$. Therefore although $\lambda^{(1)}$
is given by Eq.~(\ref{lambda_1_quant_mut}), the $\lambda'$ we use for
the $M=3,\ (n=2)$ problem is actually $\lambda' = l_{1} ( 2R_{2} +
l_{1} -1)/2$. For the general case we will see the $\lambda'$ will
have the general structure
\begin{equation}
\lambda' = \frac{1}{2} L \left( 2R_{2} + L -1 \right)\,,
\label{lambda_prime_mut}
\end{equation}
where $L$ is an integer. So having obtained the solution for 
$u_{2},\ldots,u_{M}$ from the $M-1$ solution, we need to solve the $i=1$ 
equation (\ref{psi_eqn_mut}) where $\gamma_{1}$ is a solution of  
Eq.~(\ref{gammai_choice}), that is, 
\begin{equation}
\gamma_{1}\left( \gamma_{1}+1+2m_{1}-2R_{1}\right) = 
L \left( 2R_{2} + L - 1 \right)\,.
\label{gamma_eqm_mut}
\end{equation}
This equation has two solutions: $\gamma_{1} = - L$ and 
$\gamma_{1} = 2R_{2} - 1 - L$, and since both give the same result we take 
the former. The implementation of the boundary conditions is carried out in 
the same way as in the two allele case. This is discussed in Appendix C where 
it is shown that this implies that 
\begin{equation}
\lambda^{(n)} = \frac{1}{2} \tilde{L}_{n} 
\left( 2R_{1} + \tilde{L}_{n} -1 \right)\,,
\label{lambda_n_mut}
\end{equation}
where 
\begin{equation}
\tilde{L}_{i} \equiv \sum_{j=1}^i l_{j}\,, \ \ \ R_{1} = 
\sum^{M}_{j=1} m_{j}\,. 
\label{tilde_capL_i}
\end{equation}
Here the $l_{j}$ are non-negative integers. The solutions are given explicitly 
by equations (\ref{soln_M_first_m})-(\ref{soln_M_sixth_m}).

\section{Derivation of other results}
\label{other}

In the study of most stochastic systems, the quantities which are of
interest and which are most easily calculated are the mean and
variance of the distribution and also the stationary pdf which the pdf
tends to at large times. This is still true in the problems we are
considering in this paper, where these quantities can, in general, be
rather easily obtained. Many are already known and have been given in
Section \ref{calculations} and their derivation will be briefly
discussed in this section.

We draw particular attention to the added subtleties that exist when
no mutation is allowed. In this case the exact solutions obtained so
far only hold in the \textit{open} interval $u_{i} \in (0,1)$ and do
not include the boundaries $u_{i}=0$ or $u_{i}=1$.  It is clear that
with increasing time various alleles will become extinct and the
stationary pdf will be concentrated entirely on the boundaries where
it will accumulate.  This is somewhat different to the usual picture
of absorbing states where the probability is removed entirely from the
system.  In order to calculate moments of allele frequencies, for
example, one must take care to add in the contributions from the
boundaries to those obtained by averaging the frequencies over the
pdfs we have so far determined.  This procedure will be demonstrated
later in this section.  First, however, we calculate a few quantities
of interest when the absence of a mutation process admits the fixation
and extinction of alleles.

\subsection{Fixation and extinction}
\label{other2}

\underline{Fixation.}  The probability that one allele has fixed by a
time $t$ was first obtained by Kimura \cite{kim55b} by way of a moment
formula for the distribution.  Here we derive this quantity directly
from the explicit solution when there are only two alleles. We have
described in Section~\ref{fixation} how this can be extended to any
number of alleles.

The definition of the current (\ref{current_defn}) can be written as
\begin{equation}
J(x, t) = - \frac{1}{2} \frac{\partial }{\partial x} 
\left[ x(1-x) P(x, t) \right] 
= - \frac{1}{2} (1-2x) P(x, t) - \frac{1}{2} x(1-x) 
\frac{\partial P}{\partial x} 
\end{equation}
in which the function $P(x,t)$ is the probability distribution
\emph{excluding} any boundary contributions.  We find then at the
boundary points $x=0$ and $x=1$,
\begin{equation}
\label{flux_2}
J(0, t) = - \frac{1}{2} P(0, t) \quad \mbox{and} \quad J(1, t) = \frac{1}{2}
P(1,t)\,.
\end{equation}
We then find the probability for allele $a_1$ to have fixed by time
$t$ is the sum of the current at the boundary:
\begin{equation}
f_1(x_0, t) = \int_0^t J(1,t) \D t= \frac{1}{2} \int_0^t P(1,t) \D t
\label{f_and_P}
\end{equation}
which can be evaluated by inserting the explicit expression
(\ref{soln_2}).  This procedure yields
\begin{equation}
f_1(x_0,t) = x_{0}(1-x_{0}) \sum^{\infty}_{l=0} \frac{2l+3}{l+1} (-1)^{l}
P^{(1,1)}_{l} (1-2x_{0}) \left\{ 1 - e^{-(l+1)(l+2) t/2 }
\right\}\;.
\label{Pi_xequals1_2}
\end{equation}

Using the identities (\ref{ident1}) and (\ref{ident2}) given in
Appendix C this can be written
\begin{equation}
f_1(x_0,t) = x_{0} - \frac{1}{2} \sum^{\infty}_{l=0} (-1)^{l} 
\left[ P_{l} (1-2x_0)- P_{l+2} (1-2x_0) \right] e^{-(l+1)(l+2)t/2}\,
\label{Pi_xequals1_2_alt3}
\end{equation}
where $P_l(z)$ is a Legendre polynomial, in accordance with the
result of \cite{kim55b}.  The probability that $a_1$ has been
eliminated by time $t$ is the same as the probability that $a_2$ (with
initial proportion $1-x_0$) has become fixed. In other words,
\begin{eqnarray}
f_2(x_0, t) &=& f_1(1-x_0, t) \\
&=& (1-x_0)  - \frac{1}{2} \sum^{\infty}_{l=0} 
\left[ P_{l} (1-2x_0)- P_{l+2} (1-2x_0) \right] e^{- (l+1)(l+2) t/2}\,,
\label{Pi_xequals0_2_alt1}
\end{eqnarray}
where we have used the fact that $P_{l}(-z)=(-1)^{l}P_{l}(z)$.

\medskip

\underline{Coexistence probability.} As noted in
Section~\ref{calculations}, combinatorial arguments can be used to
calculate the probability that exactly $r$ alleles coexist at time $t$
from the fixation probability just derived.  To do this, divide the
complete set of $M$ alleles into two groups $X$ and $Y$, the first
containing a particular subset of $r$ alleles.  As argued in
Section~\ref{calculations}, the probability that all of the $M-r$
alleles in $Y$ have become extinct by time $t$, leaving \emph{some} of
those in $X$ remaining is just $f_1(x_{0},t)$, where $x_0$ is the total
initial frequency of all alleles in $X$.

To find the probability that \emph{all} the alleles in $X$ continue to
coexist at time $t$, we must subtract from $f_1(x_{0},t)$ the probability
that one or more of them has been eliminated.  For example the
probability that only the pair of alleles $a_i$ and $a_j$ coexist at
time $t$ is
\[
\chi (i,j) = f_1(x_{i,0}+x_{j,0},t)-f_1(x_{i,0},t)-f_1(x_{j,0},t) \;, 
\]
as shown by Kimura \cite{kim55b}.  One then finds the probability that
exactly two alleles remain at time $t$ to be the sum of the previous
expression over all distinct pairs of indices $i,j$. This gives
\begin{equation}
\Omega_2(\vec{x}_0,t) = \sum_{i < j} f_1(x_{i,0}+x_{j,0},t) - (M-1) \sum_{i}
f_1(x_{i,0},t) \;.
\label{any_2}
\end{equation}
Similarly, the probability that only the triple of alleles $a_{i_1}$, $a_{i_2}$
and $a_{i_3}$ all coexist at time $t$ is the probability 
$f_1(x_{i_1,0}+x_{i_2,0}+x_{i_3,0},t)$ that some subset of these alleles 
remains at time $t$, minus the probability that only any particular pair or 
single allele from $a_{i_1},a_{i_2}$ and $a_{i_3}$ coexist, that is,
\begin{multline}
\chi (i_1,i_2,i_3) = f_1(x_{i_1,0}+x_{i_2,0}+x_{i_3,0}) - 
[f_1(x_{i_1,0}+x_{i_2,0},t)-f_1(x_{i_1,0},t)-f_1(x_{i_2,0},t)] \\
-[f_1(x_{i_2,0}+x_{i_3,0},t)-f_1(x_{i_2,0},t)-f_1(x_{i_3,0},t)] \\
-[f_1(x_{i_1,0}+x_{i_3,0},t)-f_1(x_{i_1,0},t)-f_1(x_{i_3,0},t)] \\
-f_1(x_{i_1,0},t)-f_1(x_{i_2,0},t)-f_1(x_{i_3,0},t)\,. 
\label{3fs}
\end{multline}
Simplifying, we find that
\begin{equation}
\chi (i_1,i_2,i_3) = - \sum^{3}_{s=1} (-1)^{s} 
\sum_{j \in \sigma_{s} (i_1,i_2,i_3)} 
f_{1} (x_{j_{1},0}+\cdots+x_{j_{s},0},t)\,,
\label{specific_comb_3}
\end{equation}
where $\sigma_s$ is a subset of $i_1,i_2,i_3$ with $s$ elements and 
$j_1,\ldots,j_s$ are the elements of that subset. Proceeding in this way, one 
finds that the probability that the $r$ alleles $a_{i_1},\ldots,a_{i_r}$ all 
coexist at time $t$ is given by
\begin{equation}
\chi (i_1,\ldots,i_r) = \sum^{r}_{s=1} (-1)^{r-s} 
\sum_{j \in \sigma_{s} (i_1,\ldots,i_r)} 
f_{1} (x_{j_{1},0}+\cdots+x_{j_{s},0},t)\,,
\label{specific_comb_r}
\end{equation}
where now the second summation is over a subset of $i_1,\ldots,i_r$ with $s$ 
elements. The result (\ref{specific_comb_r}) can be proved by induction, by
beginning with the expression for $\chi (i_1,\ldots,i_{r+1})$ (compare with 
Eq.~(\ref{3fs}) which is the case $r=2$):
\begin{equation}
\chi (i_1,\ldots,i_{r+1}) = f_{1} (x_{i_1,0}+\cdots+x_{r+1,0},t)
- \sum^{r}_{k=1} \sum_{j \in \sigma_{k} (i_1,\ldots,i_{r+1})} 
\chi (j_1,\ldots,j_k)\,.
\label{chi_r+1}
\end{equation}
If we now assume the result (\ref{specific_comb_r}) up to and
including $r$ alleles, then by substituting these expressions for
$\chi (j_1,\ldots,j_k)$ into the right-hand side of
Eq.~(\ref{chi_r+1}), we obtain the expression (\ref{specific_comb_r}),
but with $r$ replaced by $r+1$. Since we have explicitly verified the
result for low values of $r$, the result is proved.  During the course
of the proof the double summation is rearranged using $\sum^{r}_{k=1}
\sum^{k}_{s=1} \to \sum^{r}_{s=1} \sum^{r}_{k=s}$ and the fact that a
particular set $(l_1,\ldots,l_s)$ appears $\binom{r+1-s}{k-s}$ times
in the sum is used.

The analogue of Eq.~(\ref{any_2}) in this case, namely the
probability that exactly $r$ alleles remain at time $t$ is the sum of 
Eq.~(\ref{specific_comb_r}) over all distinct $r$-tuplets of indices
$i_{1},\ldots,i_{r}$:
\begin{equation}
\Omega_r(\vec{x}_0,t) = \sum_{i \in \sigma_{r} (i_1,\ldots,i_M)} 
\chi (i_1,\ldots,i_r)\,.
\label{any_r}
\end{equation}
Substituting the result (\ref{specific_comb_r}) into this equation, and using
exactly the same manipulations that were required in the inductive proof of
(\ref{specific_comb_r}), gives the general result for $r$ alleles stated 
earlier: Eq.~(\ref{prob_exactly_r}); see also \cite{lit75b}.

\medskip

\underline{Mean time to the $r$th extinction.}  As is well-known (see,
e.g., \cite{ewe79,gar04}) the mean time $\tau(\vec{x}_0)$ to reach any
boundary from an initial position $\vec{x}_0$
\begin{equation}
\bo \tau(\vec{x}_0) = -1 
\end{equation}
where $\bo$ is the backward Kolmogorov operator appearing in
(\ref{BKE}).  The boundary conditions on $\tau(\vec{x}_0)$ are that it
vanish for any $\vec{x}_0$ corresponding to a boundary point.  For the
case $M=2$ and no mutations, we seek the solution of
\begin{equation}
\frac{1}{2} x_0(1-x_0) \frac{\D^2}{\D x_0^2} \tau(x_0) = - 1 
\end{equation}
that has $\tau(0) = \tau(1) = 0$.  Two successive integration steps
yield the required answer (\ref{tau_1}).

A related quantity, which is obtained in a similar fashion, is the
mean time to fixation of allele $a_1$, \emph{given} that it does
become fixed.  This is given by
\begin{equation}
\label{fixtime}
\tau^\ast(x_0) = \frac{\int_0^\infty \D t \, t \, \frac{\D}{\D t} f_1(x_0,t)}
{\int_0^\infty \D t \, \frac{\D}{\D t} f_1(x_0,t)}\,,
\end{equation}
since $f_1(x_0,t)$ is the probability that allele $a_1$ has become fixed by
time $t$. Since $f_1(x_0,t) \to x_0$ as $t \to \infty$, the denominator 
equals $x_0$. To find the numerator, we multiply the backward Kolmogorov 
equation by $t$ and integrate over all $t$. Use of Eqs.~(\ref{flux_2}) and 
(\ref{f_and_P}) then shows that the numerator obeys the equation
\begin{equation}
\frac{1}{2} x_0(1-x_0) \frac{\D^2}{\D x_0^2} \int_0^\infty \D t \, t \, 
\frac{\D}{\D t} f_1(x_0,t) = - x_{0}\,.
\end{equation}
This yields the result
\begin{equation}
\label{fixtime_result}
\tau^\ast(x_0) = - \frac{2\left( 1-x_{0} \right)\ln \left( 1 - x_{0} \right)}
{x_0}\,.
\end{equation}

To find the mean time to the $r$th extinction event from an initial
condition with $M$ alleles, note that the probability $Q^{<}_m(M,t)$
that $m$ or fewer alleles coexist at a time $t$ can be found by
summing (\ref{prob_exactly_r}) appropriately.  One finds
\begin{eqnarray}
Q^{<}_n(M,t) &=& \sum^{m}_{r=1} \Omega_{r} (t|\und{x}_{0}) \nonumber \\ 
&=& \sum_{r=1}^{m} \sum_{s=1}^{r} (-1)^{r-s}\binom{M-s}{r-s} 
\sum f_1(x_{i_1,0} + \cdots + x_{i_s,0}, t) \nonumber \\
&=& \sum_{s=1}^{m} (-1)^{m-s} \binom{M-s-1}{m-s} \sum f_1(x_{i_1,0} + \cdots +
x_{i_s,0}, t)\,,
\end{eqnarray}
where we have rearranged the double summation using
$\sum^{m}_{k=1} \sum^{k}_{s=1} \to \sum^{m}_{s=1} \sum^{m}_{k=s}$ and also used
the identity $\sum_{k=0}^{K} (-1)^k \binom{N}{k} = (-1)^K \binom{N-1}{K}$.

Differentiating $Q^{<}_m(M,t)$ with respect to time gives the
probability that a state comprising $m$ alleles is entered during the
time interval $[t,t+\D t]$.  This corresponds to the $r$th extinction
event, where $r = M-m$.  Hence the mean time to this extinction, the
first moment of the distribution, is given by
\begin{equation}
\tau_r(M) = \sum_{s=1}^{M-r} (-1)^{M-r-s} \binom{M-s-1}{M-r-s} \sum
\int_0^{\infty} \D t \, t \frac{\D}{\D t} f_1(x_{i_1,0} + \cdots + 
x_{i_s,0}, t)\,,
\label{tau_r(M)}
\end{equation}
the denominator being unity, since $\lim_{t \to \infty} Q^{<}_m(M,t) = 1$.
However, from Eqs.~(\ref{fixtime}) and (\ref{fixtime_result}),
\begin{equation}
\int_0^\infty \D t \, t \, \frac{\D}{\D t} f_1(x_0,t) = 
-2 \left( 1 - x_{0} \right)\ln \left( 1 - x_{0} \right)\,.
\end{equation}
This gives an expression for the function which we wish to evaluate in 
Eq.~(\ref{tau_r(M)}). Changing the summation variable from $s$ to $M-s$, and
so identifying $1-x_{0}$ with $x_{i_1,0} + \cdots + x_{i_s,0}$ gives
\begin{equation}
\tau_r (M) = -2 \sum_{s=r}^{M-1} (-1)^{s-r} \binom{s-1}{r-1} \left[x_{i_1,0} +
\cdots + x_{i_s,0}\right] \ln \left[ x_{i_1,0} + \cdots + x_{i_s,0} \right]\,.
\end{equation}

\medskip

\underline{Probability of a particular sequence of extinctions.}  Let
$Q_{M,M-1,...,2}(\und{x}_0)$ be the probability that in the evolution
allele $M$ becomes extinct first, followed by allele $M-1$, $M-2$ and
so on, ending with fixation of allele $1$.  Littler \cite{lit78a}
found the result given in Eq.~(\ref{QP}), but we give a derivation
here.  We define $Q_{M,...,2}(t|\und{x}_0, 0)$ as the probability
that this sequence of extinctions has occurred by time $t$. So for
example, in the biallelic system ($M=2$), $Q_2 (t|x_0, 0) = f_1 (x_0, t)$.
Just as one can show that $f_1$ obeys the $M=2$ backward Kolmogorov equation
(by setting $x$ equal to its value on the boundary $x=1$ and relating
$f_1$ to $P(1,t|x_0,0)$), then in general one can show that 
$Q_{M,...,2}(t|\vec{x}_{0}, t_0)$ satisfies the backward Kolmogorov
 equation (\ref{BKE})
\begin{equation}
\label{QFP}
\frac{\partial}{\partial t} Q_{M,...,2}(t|\und{x}_0,0) = \bo
Q_{M,...,2}(t|\und{x}_0,0) \;.
\end{equation}

The function $Q_{M,M-1,...,2}(\und{x}_0)$ is the stationary (i.e., 
$t \to \infty$) solution of this equation that satisfies the following boundary
conditions. First, $Q_{M,...,2}(\und{x}_0, 0)=0$ for any $\vec{x}_0$
that corresponds to any allele other than $M$ already being extinct.
That is, for any $\vec{x}_0$ that has $x_{i,0}=0$ for any $i < M$. On
the hyperplane $x_{M,0}=0$, we must have that $Q_{M,...,2}(\und{x}_0,
0)$ equal to the probability of the subsequent extinctions taking
place in the desired order by time $t$.  Taking the limit
$t\to\infty$, this boundary condition implies
\begin{equation}
\label{recur}
Q_{M,...,2}(x_{1,0}, \ldots, x_{M-1,0}, 0) = 
Q_{M-1,M-2,...,2}(x_{1,0}, \ldots, x_{M-1,0}) \;.
\end{equation}
We have chosen this particular order of extinctions to demonstrate the
result as it corresponds to the ordering implied in our definition of
$\und{u}$.  In the $\und{u}$ variables, the backward Kolmogorov operator
reads
\begin{equation}
\bo = \sum_{i=1}^{M-1} \frac{u_{i,0}(1-u_{i,0})}{\prod_{j<i}(1-u_{j,0})}
\frac{\partial^2}{\partial u_{i,0} \partial u_{j,0}} \;.
\end{equation}
We conjecture a solution
\begin{equation}\label{QM_u}
Q_{M,M-1,...,2}(\und{x}_0)=u_{1,0}u_{2,0}...u_{M-1,0}\,.
\end{equation}
Clearly, $\bo$ annihilates this product, so it is a stationary
solution of (\ref{QFP}).  The boundary condition that
$Q_{M,M-1,...,2}(\vec{x}_0) = 0$ for $x_{i,0} = 0$, $i<M$ is also
obviously satisfied. Furthermore, if $x_{M,0} = 0$,
\begin{equation}
u_{M-1,0} = \frac{x_{M-1,0}}{1 - \sum_{j=1}^{M-2} x_{j,0}} = 1\,.
\end{equation}
Hence the recursion (\ref{recur}) is satisfied. It is easily
established that $Q_2(x_{1,0},x_{2,0}) = x_{1,0}$ by finding the stationary
solution of the backward Fokker-Planck equation (\ref{QFP}) with $M=2$
and imposing the boundary conditions $Q_2(1,0) = 1$ and $Q_2(0,1) =
0$. Therefore by induction, (\ref{QM_u}) is the solution required.
Rewriting in terms of the $x$ variables,
\begin{equation}
\label{QM}
Q_{M,M-1,...,2}(\und{x}_0)= x_{1,0} \frac{x_{2,0}}{1-x_{1,0}} 
\frac{x_{3,0}}{1-x_{1,0}-x_{2,0}} \cdots
\frac{x_{M-1,0}}{1-x_{1,0}-x_{2,0}- \cdots - x_{M-2,0}} \;.
\end{equation}

The probability for an arbitrary sequence of extinctions: allele $i_1$
to go extinct first, followed by $i_2, i_3, \ldots, i_{M-1}$ leaving
only allele $i_M$ can be determined by permuting the indices in
(\ref{QM}).  This then gives us (\ref{QP}).

\subsection{Stationary distribution}

We earlier obtained the time-dependent pdf valid when mutation rates
are nonzero by imposing reflecting boundary conditions.  These have the
effect of preventing currents at the boundaries, which in turn ensure
that the limiting $t\to\infty$ solution is nontrivial and hence the
complete stationary distribution.  One can check that this is the case
from (\ref{soln_M_first_m})--(\ref{soln_M_sixth_m}).  When all the
integers $l_i$ are zero, the eigenvalue in (\ref{lambda_and_L_m}) is
also zero, indicating a stationary solution.  The remaining
eigenvalues are all positive, which relate to exponentially decaying
contributions to the pdf. Retaining then only the term with
$\lambda=0$ in (\ref{soln_M_first_m}) we find immediately that
\begin{equation}
P^\ast(\und{u}) = \prod_{i=1}^{M-1} \frac{\Gamma(2R_i)}{\Gamma(2m_i)
\Gamma(2R_{i+1})}u_i^{2m_i-1}(1-u_i)^{2R_{i+1}-1}\,.
\end{equation}
It is easy to check that this distribution is properly normalised over
the hypercube $0 \le u_i \le 1$, $i=1, \ldots, M-1$.  To change
variable back to the allele frequencies $x_i$ we use the
transformation (\ref{trans_prob_M}), and using the fact that $R_i=m_i
+ R_{i+1}$, $R_M = m_M$ and $x_M = 1 - \sum_{i=1}^{M} x_i$, we arrive
at the result quoted earlier, (\ref{Pstarmut}).

When the mutation process is suppressed, one finds from
(\ref{lambda_and_L}) that all the eigenvalues are nonzero, indicating
that the distribution we have derived is zero everywhere in the limit
$t \to \infty$.  This means that the stationary solution comprises the
accumulation of probability at boundary points, as stated in
Section~\ref{calculations}.

\subsection{Moments of the distribution}

One way to obtain the moments is to perform averages over the
distribution.  The mean and variance when mutation is present can
easily be calculated in this way from (\ref{soln_2_m}).  Calculating
the mean directly from the explicit formula for the probability
distribution (\ref{soln_2}) when fixation can occur is tricky because
one must include the full, time-dependent formula for the fixation
probability at the right boundary and the integrals one has to
evaluate are not particularly convenient.

Alternatively, we can exploit the Kolmogorov equation written in the
form of a continuity equation (\ref{continuity}). This leads to a
differential equation for each moment which depends on lower
moments. The first few moments can then be found iteratively in a
relatively straightforward way.

We demonstrate the method by giving the derivation of the moments of
the distribution when two alleles are initially present. We then show
that this method extends in a straightforward way to the case when $M$
alleles are present.

Specifically for $M=2$ we have
\begin{equation}
\label{continuity_M2}
\frac{\partial P(x,t)}{\partial t} + \frac{\partial J(x,t)}{\partial
  x} = 0
\end{equation}
where the current
\begin{equation}
\label{current_M2}
J(x,t) = (m_1 - R x) P(x,t) - \frac{1}{2} \frac{\partial}{\partial
  x} x(1-x) P(x,t)
\end{equation}
and where $R=m_1+m_2$.

When mutations in either direction are present (i.e., both $m_1>0$ and
$m_2>0$) the mean of $x^k$ is given by the expression
\begin{equation}
\langle x^k(t) \rangle = \int_{0}^{1}  x^k P(x, t| x_0, 0) \D x
\end{equation}
and so
\begin{eqnarray}
\label{meanxbyparts}
\frac{\partial}{\partial t} \int_{0}^{1} x^k P(x, t | x_0, 0) \D x &=&
- \int_{0}^{1} x^k \frac{\partial}{\partial x} J(x,t) \D x  \nonumber\\ &=&
k\int_{0}^{1}x^{k-1} J(x,t) \D x - \left[ x^k J(x,t) \right]_{x=0}^{1}
\end{eqnarray}
where we have used the continuity equation (\ref{continuity_M2}) and
integrated by parts.  We have already stated that there is no current
of probability through the boundaries.  In other words, here
\begin{equation}
J(0,t) = J(1,t) = 0
\end{equation}
and so the last term in (\ref{meanxbyparts}) is zero.

When fixation is a possibility, one \emph{does} have a current at the
boundaries and, the probability that allele $a_1$ is fixed at time $t$
being $f_1(x_0, t)$.  Since the function $P(x,t|x_0,0)$ excludes
contributions from the boundary in this case, we must add these
explicitly into the mean of $x^k$.  That is,
\begin{equation}
\langle x^{k} (t) \rangle = 0 \cdot f_2(x_0, t) + \int_{0}^{1} x^k P(x,t |
x_0, 0) \D x + 1 \cdot f_1(x_0, t) \;.
\end{equation}
Taking the derivative of this expression and carrying out the integration by
parts as in (\ref{meanxbyparts}), we obtain
\begin{equation}
\frac{\partial \langle x^k(t) \rangle}{\partial t} 
= k\int_{0}^{1}x^{k-1} J(x,t) \D x  - [x^k J(x,t)]_{x=0}^{1} 
+ \frac{\partial f_1(x_0, t)}{\partial t} \;.
\end{equation}
These last two terms cancel, and so in either case we are left with
\begin{equation}
\frac{\partial \langle x^k(t) \rangle}{\partial t} = 
k\int_{0}^{1} x^{k-1}J(x,t) \D x\;.
\end{equation}

Inserting the expression (\ref{current_M2}) for the current we find
\begin{equation}
\label{momentx}
\frac{\partial \langle x^k(t) \rangle}{\partial t} = k \left[ \left(
  m_i + \frac{k-1}{2} \right) \langle x^{k-1}(t) \rangle - \left( R +
  \frac{k-1}{2} \right) \langle x^k(t) \rangle \right] \;.
\end{equation}
This reveals that moments of the distribution can be calculated
iteratively. This equation is valid whether or not the $m_i$ are
non-zero.  The equation for the mean ($k=1$) can be solved directly,
and the results used to find $\langle x^2\rangle$ ($k=2$) and so on.

When there are more than 2 alleles, a similar derivation gives the
equation for the general moment $\langle
x_1^{k_1}x_2^{k_2}...x_{M-1}^{k_{M-1}}\rangle$:
\begin{multline}
\frac{\partial \langle \prod_i x_i^{k_i}\rangle}{\partial t} = 
\sum_ik_i\left\{[m_i+\frac{1}{2}(k_i-1)]\langle x_i^{k_i-1}
\prod_{j\neq i}x_j^{k_j}\rangle - \right.\\ \left.
[\frac{1}{2}(\sum_lk_l-1)+R]\langle \prod_i x_i^{k_i}\rangle \right\}\,.
\end{multline}
Thus, again we can calculate any moment by iteration.  For example
$\langle x_ix_j \rangle$, with $i\neq j$, obeys the equation
\begin{equation}
\frac{\partial \langle x_ix_j\rangle}{\partial t} = 
m_j\langle x_i \rangle + m_i\langle x_j \rangle - 
(2R+1)\langle x_ix_j\rangle\,.
\end{equation}
and using (\ref{meanx_M}) we find the result given in
(\ref{covarx_M}).

\section{Discussion}
\label{discuss}
  
In the last decade or two the ideas and concepts of population
genetics have been carried over to a number of different fields:
optimisation \cite{mit96}, economics \cite{wit93}, language change 
\cite{chr03}, among others. While the essentials of the subject such as
the existence of alleles, and their mutation, selection and drift are
usually present in these novel applications, other aspects may not
have analogues in population genetics. Furthermore, phenomena such as
epistasis, linkage disequilibrium, the relationship between phenotypes
and genotypes, which form a large part of the subject of population
genetics, may be unimportant or irrelevant in these applications. Our
motivation for the work presented here has its roots in the
mathematical modelling of a model of language change where it is quite
plausible that the number of variants (alleles) is much larger than
two.  It was this which led us to systematically analyse the diffusion
approximation applied to multi-allelic processes, having noted that
many of the treatments in the context of biological evolution tend to
be couched in terms of a pair of alleles, the ``wild type'' and the
``mutant''.

In this paper we have shown how the Kolmogorov equation describing the
stochastic process of genetic drift or the dynamics of mutation at a
single locus may be solved exactly by separation of variables in which
the number of alleles is arbitrary. The key to finding separable
solutions is, of course, to find a transformation to a coordinate
system where the equation is separable.  The change of variable
(\ref{cofv}) we used is similar, but slightly different to one
suggested by Kimura \cite{kim56} which he showed achieved separability
up to and including $4$ alleles. Kimura was of the opinion that novel
mathematical techniques would be needed to proceed to the general case
of $M$ alleles. What we have shown here is that with our change of
variables this generalisation is possible without the need to invoke
any more mathematical machinery than was required in the standard
textbook case of $2$ alleles. A large part of the reason for this is
the way that the problem with $(M+1)$ alleles can be constructed in a
straightforward way from the problem with $M$ alleles and that with
$2$ alleles. This simple ``nesting'' of the $M$-allele problem in the
$(M+1)$-allele one is responsible for many of the simplifications in
the analysis and is at the heart of why the solution in the general
case can be relatively simply presented.

An illustration of the simple structure of this nesting is the fact
that the general solutions, valid for an arbitrary number of alleles,
with or without mutation, is made up of products of polynomials---more
specifically Jacobi polynomials. Although the higher order versions of
these polynomials can be quite complex, even after relatively short
times only those characterised by small integers are important. As
mentioned in Section \ref{u_variables}, we have given the solutions in
terms of the transformed variables, but their form in terms of the
original variables of the problem can be found using Eqs.~(\ref{cofv})
and (\ref{trans_prob_M}).  We have also presented the derivations of 
many other quantities of interest. In the interests of conciseness
we have given only a flavour of some of these: some are new, some have
already been derived by other means and yet others are simple
generalisations of the two-allele results. Yet other results are more
naturally studied in the context of the backward Kolmogorov equation,
and we also took the opportunity to gather together the most
significant, but nevertheless little-known, ones here.

We have thus provided a rather complete description of genetic drift
and mutation at a single locus. Nevertheless, there are some
outstanding questions. For example, as discussed in Appendix B, the
transformation (\ref{cofv}) does not render the Kolmogorov equation
separable for an arbitrary mutation matrix $m_{ij}$, only one where
that rate of mutation of the $A_j$ alleles to $A_i$ alleles ($i \neq
j$) occurs at a rate independent of $j$. This is the reason for this
simplified choice for the mutation matrix---a choice also made in
all other research we are aware of.  It might seem possible in
principle to find another transformation which would allow a different
form of $m_{ij}$ to be studied, but this seems difficult for a number
of reasons. For instance, the transformation must still ensure that
the diffusion part of the Kolmogorov equation is separable.
Furthermore, the matrix has $(M-1)^{2}$ entries, and the
transformation only $(M-1)$ degrees of freedom so for only certain,
restricted, forms of $m_{ij}$ will a transformation to a separable
equation be possible. Other questions involve the study of selection
mechanisms or interactions between loci using the results here as a
basis on which to build. We are currently investigating these
questions in the context of language models, but we hope that the work
reported in this paper will encourage further investigations along
these lines among population geneticists.

\section*{Acknowledgements}

GB acknowledges the support of the New Zealand Tertiary Education
Commission.  RAB acknowledges support from the EPSRC (grant GR/R44768)
under which this work was started, and a Royal Society of Edinburgh
Personal Research Fellowship under which it has been continued.

\begin{appendix}
\section{Boundary conditions in the absence of mutations}
\label{AppA}

As we have mentioned several times in the main text, although it might
be thought that adding the possibility that mutations occur would
complicate the problem, in many ways it is the situation of pure
random mating that is the richer mathematically. So, for example, the
case with mutations has a conventional stationary pdf given by
Eq.~(\ref{Pstarmut}), whereas without mutations the stationary pdf is
a singular function which exists only on some of the boundaries. Of
course, it is clear from the nature of the system being modelled that
this has to be the case, but what interests us in this Appendix is the
nature of the boundary conditions which have to be imposed on the
eigenfunctions of the Kolmogorov equation (\ref{Kol_M}) to obtain the
correct mathematical description.
 
The nature of the boundary conditions required when fixation can occur
are discussed in the literature both from the standpoint of the
Kolmogorov equation in general \cite{fel52,fel54,gar04} and in the
specific context of genetic drift \cite{ewe79,mar77}. Here we will
describe a more direct approach, which while not so general as many of
these discussions, is easily understood and illustrates the essential
points in an explicit way. It is sufficient to discuss the
one-dimensional (two allele) case, since all the points we wish to
make can be made with reference to this problem.

The question then, can be simply put: what boundary conditions should
be imposed on Eq.~(\ref{Kol_2})?  On the one hand one might feel that
that these should be absorbing boundary conditions \cite{gar04,ris89},
because once a boundary is reached there should be no chance of
re-entering the region $0<x<1$.  On the other hand, however, the
diffusion coefficient ``naturally'' vanishes on the boundaries which
automatically guarantees absorption, so there would seem to be no need
to further impose a vanishing pdf (as an absorbing boundary condition
would require). In that case, what boundary conditions should be
imposed?  Also is it even clear, given the fact that the diffusion
coefficient $x(1-x)/2$ becomes vanishingly small as the boundaries are
approached, that the boundaries can be reached in finite time?

To address these questions let us begin with the Kolmogorov equation
(\ref{Kol_2_mut}) which includes a deterministic component $A(x)$ as
well as a random component represented by the diffusion coefficient
$D(x)$. Although we are interested in the case when $A(x)$ is not
present, it will help in the interpretation of the results if we
initially include it. We also make two notational changes---we put a
subscript $I$ on $A(x)$ which will be explained below and we will
write $D(x) = g^{2}(x)/2$. The function $g(x)$ is real, since $D(x)
\geq 0$. Therefore our starting equation is
\begin{equation}
\frac{\partial P}{\partial t} = -\frac{\partial }{\partial x} 
\left[ A_{I}(x) P \right] + \frac{1}{2} \frac{\partial^{2}}{\partial x^2} 
\left[ g^{2} (x) P \right]\,.
\label{ito}
\end{equation}
It turns out that it will more useful for our purposes to write this in the 
form
\begin{eqnarray}
\frac{\partial P}{\partial t} &=& -\frac{\partial }{\partial x} 
\left[ \left\{ A_{I}(x) - \frac{1}{2} g(x) \frac{\D g}{\D x} \right\} P \right]
+ \frac{1}{2} \frac{\partial }{\partial x} 
\left[ g(x) \frac{\partial }{\partial x} \left( g (x) P \right) \right]
\nonumber \\ \nonumber \\
&=& -\frac{\partial }{\partial x} \left[ A_{S}(x) P \right] + 
\frac{1}{2} \frac{\partial }{\partial x} 
\left[ g(x) \frac{\partial }{\partial x} \left( g (x) P \right) \right]\,.
\label{strat}
\end{eqnarray}
where
\begin{equation}
A_{S} (x) = A_{I}(x) - \frac{1}{2} g(x) \frac{\D g}{\D x}\,.
\label{ito_strat}
\end{equation}
Equations (\ref{ito}) and (\ref{strat}) are known as the Ito and Stratonovich 
forms of the Kolmogorov equation respectively \cite{gar04,ris89}. There is 
no need for us to explore the differences between these two formulations; the 
only relevant point here is that the Stratonovich form is more convenient for 
our purposes.

We now transform Eq.~(\ref{strat}) by introducing the pdf $Q(y,t|y_{0},0)$
which is a function of the new variable $y$ defined by
\begin{equation}
y = \int^{x} \frac{\D x}{g(x)}\ \ \Rightarrow \ \ Q = P \frac{\D x}{\D
  y} = P g(x)\,.
\label{x_to_y} 
\end{equation} 
The transformed equation reads
\begin{equation}
\frac{\partial Q}{\partial t} = -\frac{\partial }{\partial y} 
\left[ \hat{A} (y) Q \right] + 
\frac{1}{2} \frac{\partial^{2}Q}{\partial y^2}\,, 
\label{pdf_y}
\end{equation}
where
\begin{equation}
\hat{A} (y) = \frac{A_{S} (x)}{g(x)} = \frac{A_{I} (x)}{g(x)} - \frac{1}{2}
\frac{\D g(x)}{\D x}\,.
\label{trans_A_defn}
\end{equation}
The system with an $x$-dependent diffusion coefficient has now been 
transformed to one with a state-independent diffusion coefficient, at the cost
of adding an additional factor to the deterministic term.

\begin{figure}
\begin{center}
\includegraphics[width=0.45\linewidth]{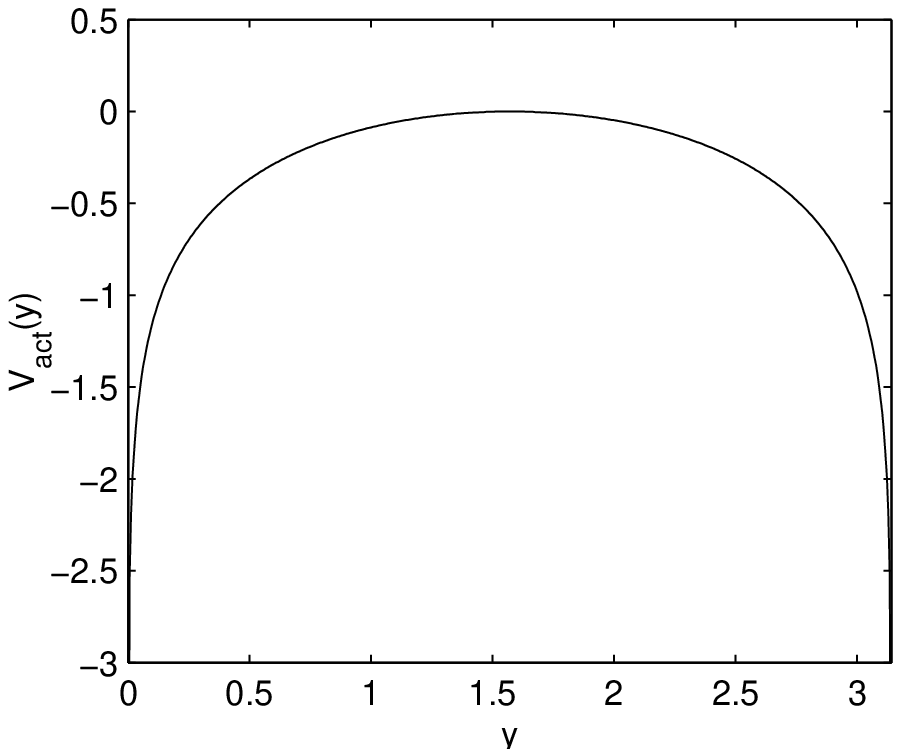}
\hspace{0.05\linewidth}
\includegraphics[width=0.45\linewidth]{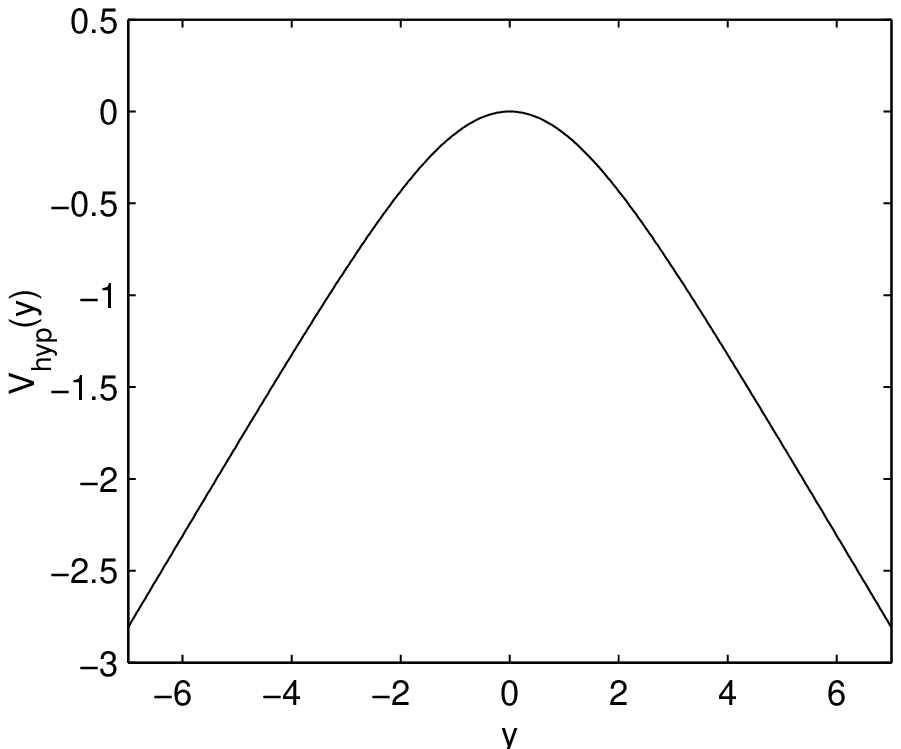}
\caption{Left: Potential $V(y)$ in the actual genetic drift
system. Right: Potential $V(y)$ in the hypothetical system described in
the text.}\label{potentials}
\end{center}
\end{figure}

The problem of interest to us has $A_{I} (x) =0$ and
$g(x)=\sqrt{x(1-x)}$. From Eq.~(\ref{x_to_y}) we find the required
transformation to be $x=\sin^{2} (y/2)$ or $\cos y = 1-2x$ with
$0<y<\pi$, and from Eq.~(\ref{trans_A_defn}) we find that $\hat{A} = -
(1/2) {\rm cot} y$. One way to understand this process intuitively is
through a mechanical analogy: if we set $\hat{A} = - \frac{\D V}{\D
y}$, then the Kolmogorov equation can be thought of as describing the
motion of an overdamped particle in the one-dimensional potential
$V(y)=-\int^{y} \hat{A} (y) \D y = (1/2) \ln \sin y$, subject to white
noise with zero mean and unit strength. An examination of
Fig.~\ref{potentials} shows that the boundaries are reached in a
finite time. More importantly, from the relation $Q=gP$ we see that
imposing the absorbing boundary conditions $Q(0, t) = Q(\pi, t) = 0$
only implies that the pdf $P$ must diverge with a power smaller than
$1/2$ at the boundaries $x=0$ and $x=1$.

These results should be contrasted with the hypothetical situation
where $A_{I} (x) = 0$, but $g(x) = x(1-x)$. In this situation we have
that $x = \left[ 1 + e^{-y} \right]^{-1}$ or $y = \ln \left( x/(1-x)
\right)$ with $-\infty < y < \infty$, and from
Eq.~(\ref{trans_A_defn}) we find that $\hat{A} = (1/2) {\rm tanh}
(y/2)$. This corresponds to a process in which an overdamped particle
is moving in the potential $V(y) = -\ln \cosh (y/2)$ and subject to
white noise. For large $|y|$, $V(y) \sim - |y|/2$, which indicates
that it will take an infinite time to reach the boundaries. This
potential is shown on the right-hand side of Fig. \ref{potentials}. 
Considerations such as these complement more mathematically rigorous 
studies by giving insights into the nature of the processes involved 
when the diffusion coefficient is dependent on the state of the system.

\section{Coordinate system in which the Kolmogorov equation is separable}
\label{appB}
\setcounter{equation}{0}

The transformation from the coordinate system defined in terms of the
set of allele frequencies, $x_i$, to that denoted by $u_i$, for which
the Kolmogorov equation is separable, is given by Eq.~(\ref{cofv}),
with the inverse transformation given by Eq.~(\ref{cofv_inverse}). In
this Appendix we explore the nature of this transformation, determine
the Jacobian of the transformation and calculate how the jump moments
in the new coordinate system are related to those in the old one. This
enables us to derive the Kolmogorov equation in the new
variables---given by Eqs.~(\ref{Kol_M_u}) and (\ref{calA_calD}) in
the main text. We begin by looking at the nature of the transformation
for low values of $M$.

\medskip

\noindent $\und{M=2}$: $0 \leq x_{1} \leq 1$. 

$u_{1}=x_{1}\ \ \Rightarrow 0 \leq u_{1} \leq 1$. 

\medskip

\begin{figure}[htb]
\begin{center}
\scalebox{0.5}{\includegraphics{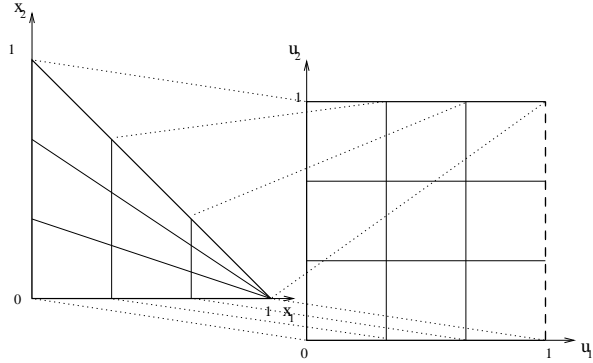}}
\caption{Schematic representation of the mapping of points in the $x_1-x_2$ 
plane onto the $u_1-u_2$ plane.}\label{cov_3}
\end{center}
\end{figure}

\noindent $\und{M=3}$: $0 \leq x_{1}, x_{2} \leq 1$. Since 
$x_{3} = 1 - x_{1} - x_{2}$ also satisfies this condition, then 
$x_{1}+x_{2} \leq 1$.  
\begin{displaymath}
u_{1} = x_{1}\,, \ \ \ u_{2} = \frac{x_2}{1-x_{1}}\,.
\end{displaymath}
Therefore, $0 \leq u_{1}, u_{2} \leq 1$, with the three lines 
$x_{1}=0\ (0 \leq x_{2} \leq 1)\,, x_{2}=0\ (0 \leq x_{1} \leq 1)$ and
$x_{1}+x_{2}=1\ (0 \leq x_{1} < 1)$ go over to the three lines 
$u_{1}=0\ (0 \leq u_{2} \leq 1)\,, u_{2}=0\ (0 \leq u_{1} \leq 1)$ and
$u_{2}=1\ (0 \leq u_{1} < 1)$ respectively. The point $x_{1}=1, x_{2}=0$ goes 
over to the line $u_{1}=1\ (0 \leq u_{2} \leq 1)$. This is illustrated in 
Fig.~\ref{cov_3}

\begin{figure}[htb]
\begin{center}
\scalebox{0.6}{\includegraphics{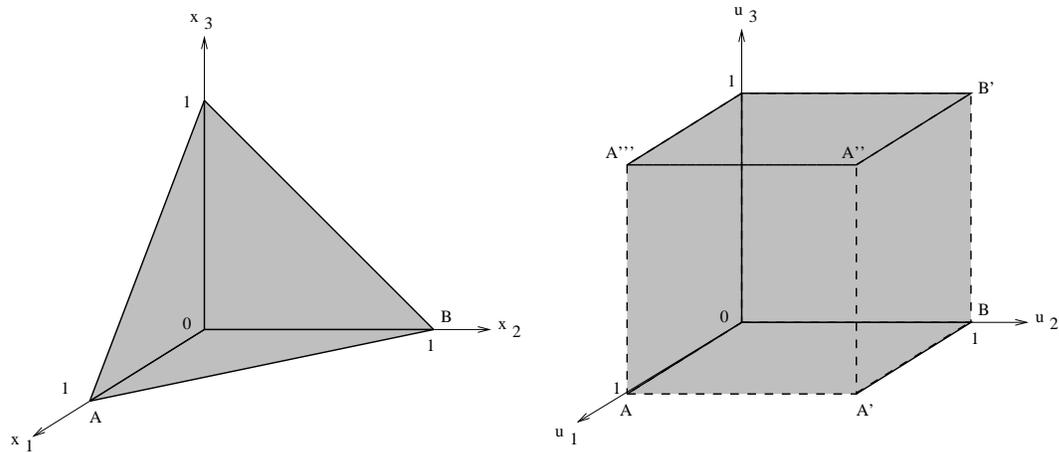}}
\caption{The range of values that can be taken by $x_i$ and $u_i$ in a 4 
allele system. The points $A, A', A'' $ and $A'''$ on the right all map onto 
the point $A$ on the left. Similarly, $B$ and $B'$ both map onto $B$.}
\label{cov_4}
\end{center}
\end{figure}

\medskip

\noindent $\und{M=4}$: $0 \leq x_{1}, x_{2}, x_{3} \leq 1$ with the additional 
condition $x_{1}+x_{2} + x_{3} \leq 1$.  
\begin{displaymath}
u_{1} = x_{1}\,, \ \ \ u_{2} = \frac{x_2}{1-x_{1}}\,, \ \ \ 
u_{3} = \frac{x_3}{1-x_{1}-x_{2}}\,. 
\end{displaymath}
Therefore, $0 \leq u_{1}, u_{2}, u_{3} \leq 1$. The planes $x_{i}=0$ get
mapped on to the planes $u_{i}=0$ ($i=1,2,3$), the plane 
$x_{1}+x_{2}+x_{3}=1,\ x_{1}+x_{2} \neq 1$ 
gets mapped on to the plane $u_{3}=1,\ 0 \leq u_{1}, u_{2} < 1$, the line
$x_{1}+x_{2}+x_{3}=1,\ x_{1}+x_{2}=1,\ x_{1} \neq 1$ gets mapped to the plane 
$u_{2}=1,\ 0 \leq u_{3} \leq 1,\ 0 \leq u_{1} < 1$ and the point    
$x_{1}+x_{2}+x_{3}=1,\ x_{1}+x_{2}=1,\ x_{1} = 1$ gets mapped to the plane 
$u_{1}=1,\ 0 \leq u_{2}, u_{3} \leq 1$. This is illustrated in 
Fig.~\ref{cov_4}.

\medskip

\noindent \underline{General $M$}: $0 \leq x_{1},\ldots, x_{M-1} \leq 1$ with 
the additional condition $x_{1}+ \ldots + x_{M-1} \leq 1$.

From Eqs.~(\ref{cofv}) and (\ref{cofv_inverse}) we see that $0 \leq
u_{i} \leq 1,\ i=1,\ldots, M-1$. The pattern from the $M=3$ and $M=4$
cases should now be clear: the hyperplanes $x_{i}=0$ get mapped on to
the hyperplanes $u_{i}=0$ ($i=1,\ldots,M-1$), the hyperplane $x_{1}+
\ldots +x_{M-1}=1,\ x_{M-1} \neq 0$, gets mapped on to the hyperplane
$u_{M-1}=1,\ 0 \leq u_{1}, \ldots, u_{M-2} < 1,\ \ldots\ $, the point
$x_{1}=1,\ x_{i}=0,\ i=2,\ldots,M-2$ gets mapped to the hyperplane
$u_{1}=1,\ 0 \leq u_{i} \leq 1,\ i=2,\ldots,M-2$.

\medskip

So, in summary, the region $0 \leq x_{i} \leq 1$ lying between the
origin and the hyperplane $x_{1}+\ldots+x_{M-1}=1$, gets mapped into
the unit hypercube $0 \leq u_{i} \leq 1$. However, parts of the
hyperplane $x_{1}+\ldots+x_{M-1}=1$ of dimensions $M-1, M-2,\ldots,1$
get mapped on to the hyperplanes $u_{i}=1,\
i=1,\ldots,M-1$. 

The Jacobian of the transformation from $\und{u}$ to $\und{x}$ is easily
found if we note from (\ref{cofv_inverse}) that
$\partial x_{i}/\partial u_{j} = 0$ if $j>i$. This implies that the Jacobian 
has zeros if the column label is greater than the row label, and so the 
determinant is just the product of the diagonal elements. Since the $i$th 
diagonal element is simply $x_{i}/u_{i}$, we have that
\begin{equation}
\frac{\partial (x_{1},\ldots,x_{M-1})}{\partial(u_{1},\ldots,u_{M-1})}=
\frac{x_1}{u_1} \frac{x_2}{u_2} \ldots \frac{x_{M-1}}{u_{M-1}} =
 \left( 1 - u_{1} \right)^{M-2} \left( 1 - u_{2} \right)^{M-3}
\ldots \left( 1 - u_{M-2} \right)\;.
\label{Jacobian}
\end{equation}
As explained in Section \ref{u_variables}, the pdfs in the $\und{u}$ variables,
denoted by $\mathcal{P}$, are related to those in the $\und{x}$ variables,
denoted by $P$, by a multiplicative factor which is simply the Jacobian:
\begin{eqnarray}
\nonumber
\mathcal{P} (\und{u}, t) &=&
P(\und{x}, t)\,\frac{\partial (x_{1},\ldots,x_{M-1})}
{\partial(u_{1},\ldots,u_{M-1})}\,, \\ \nonumber \\
\mathcal{P} (\und{u}, t_{2}; \und{v}, t_{1}) &=&
P(\und{x}, t_{2};\und{y}, t_{1})\,\frac{\partial (x_{1},\ldots,x_{M-1})}
{\partial(u_{1},\ldots,u_{M-1})}\,\frac{\partial (y_{1},\ldots,y_{M-1})}
{\partial(v_{1},\ldots,v_{M-1})}\,, \ \ldots\ \,,
\nonumber
\end{eqnarray}
from which (\ref{trans_prob}) follows.

The final point which we wish to discuss in this Appendix is the form of the
Kolmogorov equation in the new $\und{u}$ variables. As discussed in 
Section \ref{soln}, the easiest way to determine this is to relate the 
jump moments in the old coordinate system, defined through Eq.~(\ref{jm_defn}),
and those in the new coordinate system, defined through 
Eq.~(\ref{jm_defn_new}). To do this, let us begin with the first jump moment:
\begin{eqnarray}
\left\langle \Delta u_{i} (t) \right\rangle_{\und{u}(t)=\und{u}} &=& 
\int \D\und{u}' \left( u_{i}' - u_{i} \right) 
\mathcal{P} (\und{u}', t+\Delta t|\und{u}, t) \nonumber \\
&=& \int \D\und{x}' \left( f_{i} (\und{x}') - f_{i} (\und{x}) \right) 
P (\und{x}', t+\Delta t|\und{x}, t)\,,
\nonumber
\end{eqnarray}
where we have used (\ref{trans_prob}) and where $u_{i} = f_{i} (\und{x})$ is
the transformation (\ref{cofv}). Use of Taylor's theorem then gives
\begin{eqnarray}
\left\langle \Delta u_{i} (t) \right\rangle_{\und{u}(t)=\und{u}} &=& 
\sum_{j} \left\langle \Delta x_{j} (t) \right\rangle_{\und{x}(t)=\und{x}}\,
\frac{\partial f_i}{\partial x_j} \nonumber \\ 
&+& \frac{1}{2!} \sum_{j} \sum_{k} \left\langle \Delta x_{j} (t) 
\Delta x_{k} (t) \right\rangle_{\und{x}(t)=\und{x}}\, 
\frac{\partial^{2} f_i}{\partial x_j \partial x_k} + \ldots\,.
\label{trans_jm_1}
\end{eqnarray}
Since jump moments higher than the second vanish in the limit $\Delta t \to 0$,
we find from Eq.~(\ref{trans_jm_1}) that
\begin{equation}
\tilde{\alpha}_{i} (\und{u}) =  \sum_{j} \alpha_{j} (\und{x}) 
\frac{\partial u_i}{\partial x_j} + \frac{1}{2} \sum_{j} \sum_{k} 
\alpha_{jk} (\und{x}) \frac{\partial^{2} u_i}{\partial x_j \partial x_k}\,.
\label{trans_jm_1.1}
\end{equation}
Similarly,
\begin{equation}
\tilde{\alpha}_{ij} (\und{u}) =  \sum_{k} \sum_{l} \alpha_{kl} (\und{x}) 
\frac{\partial u_i}{\partial x_k}\,\frac{\partial u_j}{\partial x_l}\,.
\label{trans_jm_2}
\end{equation}
The results (\ref{trans_jm_1.1}) and (\ref{trans_jm_2}) tell us how the
jump moments transform from one coordinate system to another. In our case,
we have from Eq.~(\ref{cofv})
\begin{equation}
\label{du_by_dx}
\frac{\partial u_i}{\partial x_j} = \left\{ \begin{array}{ll} 
u^{2}_{i}/x_i , & \mbox{\ if $j < i$} \\ 
u_i/x_i , & \mbox{\ if $j=i$} \\ 
0 , & \mbox{\ if $j > i$\,,}
\end{array} \right.
\end{equation}
and
\begin{equation}
\label{d2u_by_dxdx}
\frac{\partial^{2} u_i}{\partial x_k \partial x_l} = \left\{ \begin{array}{ll} 
2 u^{3}_{i}/x_i^{2} , & \mbox{\ if $k,l < i$} \\ 
u_i^{2}/x_i^{2} , & \mbox{\ if ($k=i, l<i$) or ($l=i, k<i$)} \\ 
0 , & \mbox{\ if $k=l=i$} \\
0 , & \mbox{\ if $k>i$ or $l>i$\,.}
\end{array} \right.
\end{equation}
The key feature of the two results (\ref{du_by_dx}) and (\ref{d2u_by_dxdx}) is
that they only depend on $i$, and not on $j, k$ or $l$. This simplifies the
summations which appear in (\ref{trans_jm_1.1}) and (\ref{trans_jm_2}). A
still fairly tedious calculation shows that the second term on the 
right-hand side of (\ref{trans_jm_1.1}) is zero and that the terms with 
$i \neq j$ in (\ref{trans_jm_2}) are also zero. Specifically,
\begin{eqnarray}
\label{jm_1}
\tilde{\alpha}_{i} (\und{u}) &=&  \sum_{j} \alpha_{j} (\und{x}) 
\frac{\partial u_i}{\partial x_j} = \sum_{j} 
\left. \frac{\D x_{j}}{\D t} \right|_{\rm det}
\frac{\partial u_i}{\partial x_j} = 
\left. \frac{\D u_{i}}{\D t} \right|_{\rm det}\,. \\ \nonumber \\
\tilde{\alpha}_{ij} (\und{u}) &=& \left\{ \frac{u_i(1-u_i)}
{\prod_{k<i} (1-u_k)} \right\} \delta_{ij}\,.
\label{jm_2}
\end{eqnarray}

In this paper the only deterministic mechanism which we will consider will be
mutation, described by the differential equation
\begin{equation}
\frac{\D x_i}{\D t} = \sum_{j \neq i} m_{ij} x_{j} (t) - 
\sum_{j \neq i} m_{ji} x_{i} (t)\,,
\label{general_mut}
\end{equation}
where $m_{ij}$ is the rate of mutation of allele $j$ into allele $i$. From 
Eqs.~(\ref{du_by_dx}) and (\ref{jm_1}) 
\begin{equation}
\tilde{\alpha}_{i} = \frac{u_i}{x_i} \left\{ u_{i} \sum_{j < i}
\frac{\D x_j}{\D t} + \frac{\D x_i}{\D t} \right\}\,.
\label{explicit_firstjm}
\end{equation}
We will denote the term in the brackets in Eq.~(\ref{explicit_firstjm}) by
$\tilde{\beta}_{i} (\und{u})$, so that 
\begin{displaymath}
\tilde{\alpha}_{i} (\und{u}) = \frac{\tilde{\beta}_{i} (\und{u})}{\prod_{j<i} 
(1-u_j)}\,.
\end{displaymath}
Using Eqs.~(\ref{KM_expansion_new}) and (\ref{jm_2}), we find the 
Kolmogorov equation in the new coordinate system to be
\begin{equation}
\frac{\partial \mathcal{P}}{\partial t} = \frac{1}{\prod_{j<i} (1-u_j)}\,
\sum_{i} \left\{ - \frac{\partial }{\partial u_{i}} \left[ \tilde{\beta}_{i} 
(\und{u})\,\mathcal{P} \right] + \frac{1}{2} \frac{\partial^{2}}
{\partial u_i^2} \left[ u_{i}(1-u_i)\,\mathcal{P} \right] \right\}\,.
\label{Kol_new_gen}
\end{equation}
In Section \ref{soln} we prove that the Kolmogorov equation is separable in
the $\und{u}$ coordinate system if $\tilde{\beta}_{i} (\und{u})$ depends on
$u_{i}$ only, and not on the $u_{j},\ j \neq i$. Since $\tilde{\beta}_{i}$
involves a sum over the expression (\ref{general_mut}), which has itself to
be written in terms of the $\und{u}$ using Eq.~(\ref{cofv_inverse}), it is
clear that this will not be the case for general $m_{ij}$. However, if it
is assumed that the rate of mutation of the alleles $A_j\ (j \neq i)$ to $A_i$
occurs at a constant rate $m_i$, independent of $j$, then $m_{ij}=m_i$ for all
$j \neq i$ and (\ref{general_mut}) becomes
\begin{equation}
\frac{\D x_i}{\D t} = m_{i} \left( 1 - x_{i} (t) \right) - 
\left( \sum^{M}_{j \neq i} m_{j} \right) x_{i} (t) 
= m_{i} - \left( \sum^{M}_{j=1} m_{j} \right) x_{i} (t)\,.
\label{specific_mut}
\end{equation}
In this case
\begin{eqnarray}
\nonumber
\tilde{\beta}_{i} &=& u_{i} \sum^{i-1}_{j=1} m_{j} - R u_{i} \sum_{j<i} x_{j} 
+ m_{i} - R x_{i} \\
&=& u_{i} \sum^{i-1}_{j=1} m_{j} + R (x_{i} - u_{i}) + m_{i} - R x_{i}\,,
\nonumber
\end{eqnarray}
where $R \equiv \sum^{M}_{j=1} m_{j}$ and where we have used 
Eq.~(\ref{cofv_inverse}). We see that $\tilde{\beta}_{i} (\und{u})$ only 
depends on $u_i$ and that
\begin{equation}
\tilde{\alpha}_{i} (\und{u}) = \frac{m_{i} - R_{i} u_{i}}
{\prod_{j<i} (1-u_j)}\,,
\label{jm_1_mut}
\end{equation}
where $R_{i} \equiv \sum^{M}_{j=i} m_{j}$. This leads to the Kolmogorov 
equation
\begin{equation}
\frac{\partial \mathcal{P}}{\partial t} = \frac{1}{\prod_{j<i} (1-u_j)}\,
\sum_{i} \left\{ \frac{\partial }{\partial u_{i}} \left[ \left( R_{i} u_{i} 
- m_{i} \right)\;\mathcal{P} \right] + \frac{1}{2} \frac{\partial^{2}}
{\partial u_i^2} \left[ u_{i}(1-u_i)\;\mathcal{P} \right] \right\}\,.
\label{Kol_new_spec}
\end{equation}
This is the content of Eqs.~(\ref{Kol_M_u}) and (\ref{calA_calD}) in the main
text.

\section{Explicit solution of the second order differential equations}
\label{appC}
\setcounter{equation}{0}

In this Appendix we will provide further details relating to the solution of
the equations (\ref{psi_eqn_1}) and (\ref{psi_eqn}) subject to the 
appropriate boundary conditions. Some of the details have already been given 
in Section \ref{soln}, in particular how the substitution (\ref{f_defn}) leads
to the hypergeometric equation (\ref{hypergeo}). Consequently, much of this
Appendix will be concerned with the hypergeometric functions $F(a,b;c;u)$ 
which are the solutions of this equation. We will refer to Chapter 15 of the 
standard handbook by Abramowitz and Stegun~\cite{abr65} for the formul\ae\ 
that we use, but several appear sufficiently often that it is worth stating 
them explicitly here:
\begin{itemize}
\item[1.] If $c-a-b>0$ and $c \neq 0,-1,-2,\ldots$, then
\begin{displaymath}
F(a,b;c;1) = \frac{\Gamma (c) \Gamma (c-a-b)}{\Gamma (c-a) \Gamma (c-b)}\,.
\end{displaymath}
\item[2.] $F(a,b;c,u) = (1-u)^{(c-a-b)} F(a',b';c;u)$ where $a'=c-a$ and 
$b'=c-b$. Note that $c-a'-b'=-(c-a-b)$.

\item[3.] If $a$ or $b$ is a non-positive integer then the power series for
$F(a,b;c;u)$ terminates:
\begin{displaymath}
F(-n,\alpha + 1 +\beta +n; \alpha +1;u) = 
\frac{\Gamma (n+1) \Gamma (\alpha + 1)}{\Gamma (n + \alpha + 1)}
P^{(\alpha,\beta)}_{n} (1-2u)\,,
\end{displaymath}
where $P^{(\alpha,\beta)}_{n}$ is a Jacobi Polynomial of order $n$.
\end{itemize}
As we will see, the solution of the hypergeometric differential equation  
typically proceeds as follows. The general solution to the equation 
(\ref{hypergeo}) is written out as $f(u)=A f_{1} (u) + B f_{2} (u)$, where 
$f_{1} (u)$ and $f_{2} (u)$ are two independent solutions of the equation, 
and $A$ and $B$ are two arbitrary constants. Examining the general solution in
the vicinity of the boundary at $u=0$ rules out one of the solutions, so that,
for instance one must take $B=0$ in order to satisfy the boundary condition
there. The remaining solution is now examined in the vicinity of the other
boundary at $u=1$. If the solution has the form $F(\hat{a},\hat{b};\hat{c};u)$ 
with $\hat{c}-\hat{a}-\hat{b}>0$ then we may use Result 1 above. If 
$\hat{c}-\hat{a}-\hat{b}<0$, then we may use Result 2 followed by Result 1. 
If it is found that either $\hat{a}$ or $\hat{b}$ is a non-positive integer,
then Result 3 may be invoked.

We now separately examine the specific equations encountered in the
absence and presence of mutations.

\subsection{No mutations}  

The values of the parameters are given in Section \ref{soln_neutral}, where it
is seen from Eq.~(\ref{abc_defn_neut}) that $c=2$. One of the solutions of the
hypergeometric equation when $c=2$ (suppose it is $f_{2}$) has the following
behaviour for small $u$~\cite{abr65}:
\begin{equation}
f_{2} (u) \sim \frac{1}{(1-a)(1-b) u}\,, \ \ {\rm as} \ u \to 0\,,
\label{neut_utozero}
\end{equation}
and so this solution would not lead to a finite pdf as $u \to
0$---recall from Appendix~A that only divergences less strong than
$u^{-1/2}$ will be tolerated. Therefore $B=0$. The remaining solution
is $f_{1} (u) = F(a,b;2;u)$. From Eq.~(\ref{abc_defn_neut}),
$2-a-b=-1-2\gamma_{i}$, so if we choose $\gamma_{i}>0$, as described
in Section \ref{soln_neutral}, we may use Result 2 above to show that
\begin{displaymath}
f_{1} (u) = \left( 1 - u \right)^{-2\gamma_{i}-1} F(2-a,2-b;2;u)\,.
\end{displaymath}
Now $2-(2-a)-(2-b)=a+b-2=2\gamma_{i}+1>0$ and so we may use Result 1 to show
that 
\begin{eqnarray}
f_{1} (u) &\sim& \left( 1 - u \right)^{-2\gamma_{i}-1} 
\frac{\Gamma (2\gamma_{i} + 1)}{\Gamma (a) \Gamma (b)}\,, 
\ \ {\rm as} \ u \to 1
\nonumber \\
\Rightarrow \ \ \psi (u) &\sim& \left( 1 - u \right)^{-\gamma_{i}-1} 
\frac{\Gamma (2\gamma_{i} + 1)}{\Gamma (a) \Gamma (b)}\,, 
\ \ {\rm as} \ u \to 1\,.
\label{neut_utoone}
\end{eqnarray}
We may once again use the result of Appendix A to deduce that this term cannot 
be present since it is diverging too fast as $u \to 1$ (the analysis of 
Appendix A can be straightforwardly generalised to cover the equations that
are found from separation of the $M$ allele Kolmogorov equation). We therefore
require that either $a$ or $b$ equals $-l$, where $l$ is a non-negative 
integer. Since the hypergeometric function is symmetric in $a$ and $b$ it does 
not matter which we choose: let $a=-l$. From Result 3 this means that the 
hypergeometric function is terminating, and more specifically,
\begin{equation}
f_{1} (u) = F(-l,2\gamma_{i}+l+3;2;u) = \frac{1}{l+1} 
P^{(1,2\gamma_{i}+1)}_{l} (1-2u)\,.
\label{nuet_finalf}
\end{equation}
As shown in Section \ref{soln_neutral}, $\gamma_{i}=L_{i-1}$, if we take 
$\gamma_{i}>0$, and so up to a constant
\begin{equation}
\psi_{\lambda^{i};\lambda^{(i-1)}} (u) = \left( 1 - u \right)^{L_{i-1}}
P^{(1,2\gamma_{i}+1)}_{l} (1-2u)\,.
\label{nuet_finalpsi}
\end{equation}
We discuss some properties of Jacobi Polynomials below.

Note that if we had made the choice $\gamma_{i}<0$ in 
Section \ref{soln_neutral} (\textit{i.e.} $\gamma_{i}=-L_{i-1}-1$), then the
argument given above would still rule out solution $f_{2} (u)$ using 
Eq.~(\ref{neut_utozero}), but we would have found the remaining solution to be 
$f_{1} (u) = F(a' ,b' ;2;u)$ where $a'$ and $b'$ are the parameter solutions 
relevant for the choice $\gamma_{i} <0$ and are related to those for the 
former $\gamma_{i}>0$ choice by $a' = 2-a$ and $b' = 2-b$. This can be easily
seen from Eq.~(\ref{abc_defn_neut}): if $\gamma_{i}$ is one solution of 
$\gamma_{i}(\gamma_{i}+1) = 2\lambda'$, then a second one is 
$\gamma_{i}' = -1 -\gamma_{i}$. But then Eq.~(\ref{abc_defn_neut}) is unchanged
if $\gamma_{i}, a$ and $b$ are replaced by $\gamma_{i}' ,a' =2-a$ and 
$b' =2-b$. So for the choice $\gamma' =-L_{i-1}-1$ we have shown that 
\begin{displaymath}
\psi (u) = \left( 1 - u \right)^{-L_{i-1}-1} F(2-a,2-b;2;u) \sim
\left( 1 - u \right)^{-\gamma_{i}-1} \frac{\Gamma (2\gamma_{i} + 1)}
{\Gamma (a) \Gamma (b)}\,, \ \ {\rm as} \ u \to 1\,,
\end{displaymath}
where the $\gamma_{i}, a$ and $b$ are those for the $\gamma_{i}>0$ choice. This
is exactly Eq.~(\ref{neut_utoone}) and so we again deduce that $a=-l$, where 
$l$ is a non-negative integer. Using Result 2 we then have that 
\begin{eqnarray}
f_{1} (u) &=& \left( 1 - u \right)^{2\gamma_{i}+1} 
F(-l,2L_{i-1}+l+3;2;u) \nonumber \\
\Rightarrow \ \ \psi_{\lambda^{i};\lambda^{(i-1)}} (u) &=& 
\left( 1 - u \right)^{L_{i-1}}P^{(1,2\gamma_{i}+1)}_{l} (1-2u)\,, \nonumber
\end{eqnarray}
which is Eq.~(\ref{nuet_finalpsi}).

The final step in calculating the solutions is to apply the initial
conditions to find the appropriate coefficients. We use the
appropriate orthogonality relationship (which we shall give below) and
use the method described in
(\ref{initial_conds})-(\ref{post_initial_conds}), and find
$w(\und{u}_0)$ and $c_{l_i}(L_i)$ as stated in (\ref{soln_M_first})
and (\ref{soln_M_third}).

\subsection{Mutations present}  

The values of the parameters are given in Section \ref{soln}, where it
is seen from Eq.~(\ref{abc_defn}) that $c=2(1-m_{1})$. The nature of the
solutions of Eq.~(\ref{hypergeo}) depends on whether $c$ is an integer, and
if it is, what its value is~\cite{abr65}. The simplest case is if $c$ is not 
an integer, then the general solution has the form
\begin{equation}
A F(a,b;c;u) + B u^{1-c} F(a-c+1,b-c+1;2-c;u)\,.
\label{c_noninteger}
\end{equation}
Both of these hypergeometric functions are of the form $F(a,b;c,u)$
with $c-a-b>-1$ and so have power-series expansions near
$u=0$~\cite{abr65}.  Substituting the expression (\ref{c_noninteger})
into the boundary condition (\ref{reflectbc_f}) gives $A=0$. The
analysis when $c=1$ and $c=0,-1,-2,\ldots$ has to be done
separately. One of the solutions has a factor of $\ln u$ in these cases,
but it turns out that, while these solutions have a current which 
does not diverge at $u=0$, the current is non-zero, which is forbidden by 
the reflecting boundary conditions. Therefore this solution must 
be absent. It is found that in all cases the implementation of the boundary 
condition at $u=0$ shows that the required solution is 
$f (u) = u^{1-c} F(a-c+1,b-c+1;2-c;u)$.

We now note that the conditions which determine the parameters $a$ and $b$ 
from Eq.~(\ref{abc_defn}) can be simplified if we introduce a new parameter
$k_i$ through $a=1+\gamma_{i}+k_{i}$. Then $b=\gamma_{i}+2-2R_{i}-k_{i}$,
where $k_{i}(2R_{i}+k_{i}-1)=2\lambda$. So choosing $\gamma_{1}=-L$, this
being one of the solutions of Eq.~(\ref{gamma_eqm_mut}), we have that
$c=2(1-m_1),\ a=1+k_{1}-L$ and $b=2-2R_{1}-k_{1}-L$ and therefore
$f(u) = u^{2m_{1}-1} F(2m_{1}+k_{1}-L,1-R_{2}-k_{1}-L;2m_{1};u)$. Using Result 
2 we may write this in an alternative form:
\begin{equation}
f_{1} (u) = u^{2m_{1}-1} \left( 1 - u \right)^{2(L+R_{2})-1} 
F(L-k_{1},k_{1}+L+2R_{1}-1;2m_{1};u)\,.
\label{alt_form}
\end{equation}
We may now apply the boundary condition that the current vanishes when 
$u=1$, which leads to Eq.~(\ref{reflectbc_f}). After some algebra we find
that this condition implies  
\begin{eqnarray}
0 &=& \lim_{u \to 1} \left\{ u^{2m_1} (1-u)^{-L} \right\} 
\left\{ -L F(2m_{1}+k_{1}-L,1-R_{2}-k_{1}-L;2m_{1};u) \right. \nonumber \\
&+& \left. \frac{(L-k_{1})(2R_{1}+k_{1}+L-1)}{2m_1}\,
F(2m_{1}+k_{1}-L,1-R_{2}-k_{1}-L;2m_{1}+1;u) \right\}\,. \nonumber
\end{eqnarray}
We may now use Result 1 to show that for $L \geq 1$ the term in the second 
set of curly brackets has a finite limit as $u \to 1$, specifically this term 
tends to
\begin{displaymath}
\frac{(2R_{2}+L-1) \Gamma (2m_1) \Gamma (2R_{2}+2L-1)}
{\Gamma (L-k_{1}) \Gamma ( 2R_{1}+k_{1}+L-1)}\,.
\end{displaymath}
Thus either $L-k_{1}$ or $2R_{1}+k_{1}+L-1$ must be a non-positive integer 
if we are to obtain a finite current (let alone a zero-current) as $u \to 1$.
The case $L=0$ may be treated separately, since the first term in the bracket,
to which Result 1 does not immediately apply is absent. The same result is 
found. To show that these conditions also imply a zero current, we may 
substitute $k_{1}=L+l$ or $k_{1}=1-2R_{1}-L-l$ into Eq.~(\ref{alt_form}) and
find, when $k_{1}+L+l$ for instance that
\begin{eqnarray}
0 &=& \lim_{u \to 1} \left\{ u^{2m_1} (1-u)^{2R_{2}+L-1} \right\} 
\left\{ -L F(-l,2R_{1}+2L-1+l;2m_{1};u) \right. \nonumber \\
&+& \left. (1-u)\,\frac{\D}{\D u}\,
F(-l,2R_{1}+2L-1+l;2m_{1}+1;u) \right\}\,. 
\label{bc_uequals1}
\end{eqnarray}
Since by Result 3 the two hypergeometric functions are polynomials of order 
$l$, the right-hand side is indeed zero if $L \geq 1$, and also when $L=0$,
since the first term in the second curly bracket of Eq.~(\ref{bc_uequals1})
is absent. Note that if we had made the choice $k_{1}=1-2R_{1}-L-l$ then the
hypergeometric functions which would have appeared would have been of the form
$F(2R_{1}+l+1+l,-l;2m_{1};u)$, which again are polynomials by Result 3.

Therefore, returning to Eq.~(\ref{alt_form}), we have found that
\begin{displaymath}
f_{1} (u) = u^{2m_{1}-1} \left( 1 - u \right)^{2(L+R_{2})-1} 
F(-l,2(L+R_{1})-1+l;2m_{1};u)\,.
\end{displaymath}
Using Result 3 this is, up to a constant,
\begin{equation}
f_{1} (u) = u^{\alpha} (1-u)^{\beta} P^{(\alpha,\beta)}_{l} (1-2u)\,,
\label{penultimateform}
\end{equation}
where $\alpha = 2m_{1}-1$ and $\beta=2(R_{2}+L)-1$. Writing it in terms of
the function $\psi$ we have that
\begin{equation}
\psi^{(1)}_{\lambda^{n};\lambda^{n-1}} (u_1) = u_{1}^{2m_{1}-1}
(1-u_1)^{2R_{2}+L-1} P^{(2m_{1}-1,2R_{2}+2L-1)}_{l} (1-2u_1)\,.
\label{final_mut_form}
\end{equation}
It is also straightforward to check that if we made the choice 
$\gamma_{1}=2R_{2}+L-1$, rather than $\gamma_{1}=-L$, then the analysis near
$u=0$ would be as before, but then use of Result 2 would then show that the
result we obtain was exactly as before (but with $a$ and $b$ interchanged).

In summary, we have shown that $k_{1}=L+l$, where $l$ is a non-negative 
integer. Thus for $M=2$, when $\gamma_{1}=0\ \Rightarrow\ L=0$, and so from 
$2\lambda = k_{1}(2R_{1}+k_{1}-1)$, we obtain Eq.~(\ref{lambda_1_quant_mut}).
For $M=3$, we use Eq.~(\ref{lambda_prime_mut}), with $L=l_{1}$ and so
obtain $\lambda^{(2)}$ given by 
\begin{equation}
\lambda^{(2)} = \frac{1}{2} L' \left( 2R_{1} + L' -1 \right)\,,
\label{lambda_2_mut}
\end{equation}
where $L' = L+l$, and $l$ is a non-negative integer and where 
$R_{1} = \sum^{3}_{j=1} m_{j}$. Continuing in this way we find the result given
by Eq.~(\ref{lambda_n_mut}).

We again use the orthogonality properties of the Jacobi Polynomials as
described in (\ref{initial_conds})-(\ref{post_initial_conds}) to
define $c_{l_i}$ in (\ref{soln_M_fourth_m}).

\subsection{Jacobi Polynomials}

All the solutions for the pdfs considered in this paper consist of Jacobi
Polynomial $P^{(\alpha,\beta)}_{l} (1-2u)$. All properties of these 
functions that we require are given in Chapter 22 of Abramowitz and 
Stegun~\cite{abr65}, but here we will list the key results that we use.

The first point to note is that the natural variable for these functions is 
$z \equiv 1 - 2u$. Thus while $u \in [0,1],\ z \in [-1,1]$. They satisfy the
following orthogonality relation in the $u$ variable:
\begin{multline}
(2l+\alpha+\beta+1)\,\frac{\Gamma(l+1)\,\Gamma
(l+\alpha+\beta+1)}{\Gamma (l+\alpha+1) \Gamma(l+\beta+1)} \times\\
\int^{1}_{0} \D u\, u^{\alpha} (1-u)^{\beta}
P^{(\alpha,\beta)}_{l} (1-2u) P^{(\alpha,\beta)}_{l'} (1-2u)
=\delta_{l,l'}\,,
\label{orthog}
\end{multline}
where $\alpha,\beta>-1$. In the eigenfunctions calculated when mutations are
present, the functions $\psi^{(i)}$ which make up the right eigenfunction 
$\Phi_{\lambda}$ as defined in Eq.~(\ref{soln_M_second_m}), contain the
factor $u^{\alpha} (1-u)^{\beta}$ which appears in Eq.~(\ref{orthog}), as
can be seen in Eq.~(\ref{soln_M_third_m}). On the other hand the functions 
$\theta^{(i)}$ which make up the left eigenfunction $\Theta_{\lambda}$ as 
defined in Eq.~(\ref{soln_M_fifth_m}), do not contain this factor, as
can be seen in Eq.~(\ref{soln_M_sixth_m}). The result is that the right and
left eigenfunctions satisfy the simple orthogonality relation 
$\int \D \und{u}\,\Theta_{\lambda} (\und{u}) \Phi_{\lambda '} (\und{u}) = \delta_{\lambda,\lambda '}$.

Secondly, Kimura uses the notation of a $J$ polynomial, which relates to the 
Jacobi polynomial as follows:
\begin{equation}
J_{n}(\alpha+2,\alpha+1,1-x)=\frac{\Gamma(n+1)\Gamma(\alpha+1)}
{\Gamma(n+\alpha+1)}(-1)^nP_n^{(1,\alpha)}(1-2x)\,.
\end{equation}

In the derivation of the fixation probability In Section~\ref{fixation} we 
make use of the following relationship between Jacobi and Legendre polynomials
which can be found in Ref.~\cite{abr65}:
\begin{equation}\label{ident1}
(1-z^2)P_{l}^{(1,1)}(z)=\frac{2(l+1)}{(l+2)} \left[ P_{l}(z) - 
z P_{l+1}(z) \right]\,,
\end{equation}
as well as this identity for Legendre Polynomials, which can be found in 
Ref.~\cite{gra94}:
\begin{equation}\label{ident2}
zP_{l+1} (z) - P_{l} (z) = \frac{(l+2)}{(2l+3)}[ P_{l+2} (z) - P_{l} (z)]\;.
\end{equation}

\end{appendix}


\begin{thebibliography}{10}
\bibitem{cro05}
W.~Croft, The relevance of an evolutionary model to historical linguistics, to
  be published.
\end{thebibliography}
\end{document}